\documentclass[twoside]{article}
\usepackage{docbib}
\usepackage{graphicx} 

\newfont{\sss}{cmbxti10} 
\newfont{\kleiner}{cmbx9}
\newfont{\foot}{cmr9} 
\newfont{\ital}{cmti9}
\renewcommand{\it}{\ital}

\newfont{\Gross}{cmbx10 scaled 1400}

\newcommand{\abb}[3]{\begin{figure}[h]
\begin{center}\vspace*{-2mm} #1 \end{center} \vspace*{-3mm}
{\hfill\parbox[b]{5.5cm}{\caption[]{#2\label{#3}}}\hfill}
\end{figure}}

\renewcommand{\vec}[1]{\relax\ifmmode\mathchoice
{\mbox{\boldmath$\relax\displaystyle#1$}}
{\mbox{\boldmath$\relax\textstyle#1$}}
{\mbox{\boldmath$\relax\scriptstyle#1$}}
{\mbox{\boldmath$\relax\scriptscriptstyle#1$}}\else
\hbox{\boldmath$\relax\textstyle#1$}\fi}

\begin{document}
\thispagestyle{empty}
\pagestyle{myheadings}
\markboth{{\kleiner INTERNATIONAL JOURNAL OF CHAOS THEORY AND 
APPLICATIONS\mbox{\ \ \ }}}
{\mbox{\ \ \ \ \ } 
{\em Dirk Helbing and Tadeusz Platkowski}{\rm : Self-Organization 
Induced by Fluctuations}}

\onecolumn
\twocolumn[\protect\hsize\textwidth\columnwidth\hsize\csname @twocolumnfalse\endcsname
{\vspace*{-15mm}
\begin{center}
 {\bf INTERNATIONAL JOURNAL OF CHAOS THEORY AND APPLICATIONS\\
  Volume ? (2000), Nr. ? }\\
 \mbox{}\hrulefill\mbox{}\\[9mm]
 \,\hfill {\sss Invited Paper}\\[9mm]    
{\Gross Self-Organization in Space and Induced by Fluctuations}\\[1cm]
{\large
Dirk Helbing$^{\circ\,\ast}$ and Tadeusz Platkowski$^{\circ\,+}$}\\[1mm] 
{\normalsize\it $^\circ$ II. Institute of Theoretical Physics, University of Stuttgart,\\
\normalsize\it Pfaffenwaldring 57/III, 70550 Stuttgart, Germany\\
\normalsize\it $^\ast$ Collegium Budapest, Institute of Advanced Study, 
Szeth\'aroms\'ag u. 2, \\
\normalsize\it  H-1014 Budapest, Hungary\\
\normalsize\it $^+$Dept. of Mathematics, Informatics and Mechanics,\\ 
\normalsize\it University of Warsaw, Banacha 2, 02-097 Warsaw, Poland
}\\[6mm]
\end{center}\par
{\bf Abstract\footnotemark[1]}\\[1mm]
We present a simple discrete model for the non-linear spatial interaction of
different kinds of ``subpopulations'' composed of identical moving entities
like particles, bacteria, individuals, etc. The model allows to mimic a variety
of self-organized agglomeration and segregation phenomena. By relating it to
game-theoretical ideas, it can be applied not only to attractive and
repulsive interactions in physical and chemical systems, but also to the
much richer combinations of positive and negative interactions found in
biological and socio-economic systems. Apart from investigating symmetric
interactions related to a continuous increase of the ``overall success''
within the system (``self-optimization''), we will focus on cases, where
fluctuations further or induce self-organization, even though the initial
conditions and the interactions are assumed homogeneous in space
(translation invariant).\\[5mm]
{\bf Keywords:} {\sss Self-organization, self-optimization, game theory,
fluctuation-induced transition, agglomeration, segregation.}\\[9mm]\mbox{ }
}
]
\footnotetext[1]{Manuscript received: xxxx; accepted: xxxx}

\section{\large Introduction}

Although the biological, social, and economic world are full of
self-organization phenomena, many people believe that the dynamics behind them
is too complex to be modelled in a mathematical way. 
Reasons for this are the huge
number of interacting variables, most of which cannot be quantified, plus
the assumed freedom of decision-making or large fluctuations within
biological and socio-economic systems. However, in many situations, the
living entities making up these systems decide for some (more or less)
optimal behavior, which can make the latter describable or predictable to a
certain extend 
\cite{We91,game1,game2,game3,gamedyn,Hub,quantsoc,frank,Lewenstein,Galam}. 
This is even more the case for the behavior 
shown under certain constraints like, for
example, in pedestrian or vehicle 
dynamics \cite{book,pre,phasediag}. While
pedestrians or vehicles can move freely at small traffic densities, at large
densities the interactions with the others and with the boundaries of the
street confines them to a small spectrum of moving behaviors. Consequently,
empirical traffic dynamics can be reproduced by simulation models
surprisingly well \cite{book,pre,phasediag,solid,scatter,opus,trail}.
\par
In this connection, it is also interesting to mention some insights gained
in statistical physics and complex systems theory: Non-linearly interacting
variables do not change independently of each other, and in many cases there
is a separation of the time scales on which they evolve. This often allows
to ``forget'' about the vast number of rapidly changing variables, which are
usually determined by a small number of ``order parameters'' and treatable as
fluctuations \cite{Haken,Adv}. In the above mentioned
examples of traffic dynamics, the order parameters are the traffic density
and the average velocity of pedestrians or vehicles.
\par
Another discovery is that, by proper transformations or scaling, many
different models can by mapped onto each other, i.e. 
they behave basically the same \cite{Haken,Adv,seg3,phasediag}. 
That is, a certain class of models displays the same kinds of
states, shows the same kinds of transitions among them, and can be described
by the same ``phase diagram'', displaying the respective states as a function
of some ``control parameters'' \cite{Haken,Adv}. 
We call such a class of models a
``universality class'', since any of these models shows the same kind of
``universal'' behavior, i.e., the same phenomena. Consequently, one usually tries
to find the simplest model having the properties of the universality class.
While physicists like to call it a ``minimal model'', ``prototype model'', or
``toy model'', mathematicians named the corresponding mathematical equations
``normal forms'' \cite{manneville,Zee77,Haken,Adv}. 
\par
Universal behavior is the reason of the great
success of systems theory \cite{Ber68,Bu67,Rap86} 
in comparing phenomena in seemingly
completely different systems, like physical, biological, or social ones.
However, since these systems are composed of different entities and their
corresponding interactions can be considerably different, it is not always
easy to identify the variables and parameters behind their dynamics. It can
be helpful to take up game-theoretical ideas, here, quantifying
interactions in terms of payoffs 
\cite{game1,game2,game3,gamedyn,Hub,quantsoc,Ebeling,gd}. 
This can be applied to
positive (profitable, constructive, cooperative, symbiotic) or
negative (competitive, destructive) interactions in socio-economic or
biological systems, but to attractive and repulsive interactions in
physical systems as well 
\cite{helvic}.
\par
In the following, we will investigate a simple model for interactive motion
in space allowing to describe (i) various self-organized agglomeration
phenomena, like settlement formation, and segregation phenomena, like ghetto
formation, emerging from different kinds of interactions and (ii)
fluctuation-induced ordering or self-organization phenomena. 
\par
Noise-related phenomena can be quite surprising and have, therefore,
recently attracted the interest of many researchers. For example, we 
mention stochastic resonance \cite{stochres}, 
noise-driven motion \cite{biol1a,biol1c}, and ``freezing by heating''
\cite{freezing}. 
\par
The issue of order through fluctuations
has already a considerable history. Prigogine has discussed it
in the context of structural instability with respect to the 
appearance of a new species \cite{Pri,Pri2}, but this is not related
to the approach considered in the following. 
\par
Moreover, since both, the
initial conditions and the interaction strengths in our model are 
assumed independent of the position in space, 
the fluctuation-induced self-organization
discussed later on must be distinguished from 
so-called ``noise-induced  transitions'' as well, 
where we have a space-dependent diffusion coefficient which 
can induce a transition \cite{HoLe84}. 
\par
Although our model 
is related to diffusive processes, 
it is also different from reaction-diffusion systems
that can show fluctuation-induced self-organization phenomena known as
Turing patterns
\cite{turing,turing1,turing2,turing3,turing4,turing5}, 
which are usually periodic in space. The
noise-induced self-organization that we find seems to have (i) no typical
length scale and (ii) no attractor, since our model is
translation-invariant.
This, however,
is not yet a final conclusion and still subject to investigations. 
\par
We also point out that, in the case of spatial invariance, self-organization
directly implies spontaneous symmetry-breaking, and we expect a pronounced
history-dependence of the resulting state. Nevertheless, when averaging over a
large ensemble of simulation runs with different random seeds, we again
expect a homogeneous distribution, since this is the only result compatible
with translation invariance.
\par
Finally, we mention that our results do not fit into the concept of
noise-induced transitions from a metastable disordered 
state (local optimum) to a stable ordered
state (global optimum), 
which are, for example, found for undercooled fluids, metallic glasses, 
or some granular systems \cite{gran1,gran2,gran3}. 

\section{\large Discrete Model of Interactive Motion in Space}

Describing motion in space has the advantage that the essential variables
like positions, densities, and velocities are well measurable, which allows
to calibrate, test, and verify or falsify the model. Although we will focus
on motion in ``real'' space like the motion of pedestrians or bacteria, our
model may also be applied to changes of positions in abstract spaces, e.g. to 
opinion changes on an opinion scale \cite{quantsoc,Hel93}. 
There exist, of course,
already plenty of models for motion in space, and we can mention only a few
\cite{We91,quantsoc,book,pre,phasediag,solid,scatter,opus,trail,Haken,Adv,seg3,%
manneville,Ebeling,helvic,biol1a,biol1c,freezing,%
HoLe84,turing,turing1,turing2,turing3,turing4,%
turing5,gran2,Keizer,Hel93,granular,eshel,jacob2,biol6,biol2,%
biosys,millonas}. Most of them are,
however, rather specific for certain systems, e.g., for fluids or for
migration behavior.
\par
For simplicity, we will restrict the following considerations to a
one-dimensional space, but a generalization to higher dimensions is
straightforward. The space is divided into $I$ equal cells $i$ which can be
occupied by the entities. We will apply periodic boundary conditions, i.e.
the space can be imagined as a circle. In our model, we group the $N$
entities $\alpha$ in the system into homogeneously behaving subpopulations $%
a $. If $n_i^a(t)$ denotes the number of entities of subpopulation $a$ in
cell $i$ at time $t$, we have the relations
\begin{equation}
\sum_i n_i^a(t) = N_a, \qquad
\sum_a N_a = N. 
\end{equation}
We will assume that the numbers $N_a$ of entities belonging to the
subpopulations $a$ do not change. It is, however, easy to take additional
birth and death processes and/or transitions of individuals from one
subpopulation to another into account \cite{We91}.
\par
In order not to introduce any bias, we start our simulations with a
completely uniform distribution of the entities in each subpopulation over
the $I$ cells of the system, i.e., $n_i^a(0) = n_{\rm hom}^a 
= N_a/I$, for which we choose a
natural number. At times $t \in \{1,2,3,...\}$, we apply the following
update steps, using a random sequential update (although a parallel
update is possible as well,
which is more efficient \cite{Wolfram,Stauffer}, but normally less realistic 
\cite{Bernardo} due to the assumed synchronous updating):
\par
{\em 1st step:\/} For updating of the state of entity $\alpha$, given
it is a member of subpopulation $a$ and located in cell $i$,
determine the so-called (expected) ``success'' according to the formula
\begin{equation}
S_{a}(i,t)=\sum_{b}P_{ab}\, n_{i}^{b}(t)+\xi _{\alpha }(t) \, .
\label{formu} 
\end{equation}
Here, $P_{ab}$ is the ``payoff'' in interactions of an entity of
subpopulation $a$ with an entity of subpopulation $b$.
The payoff $P_{ab}$ is positive for attractive, profitable, constructive, or
symbiotic interactions, while it is negative for repulsive, competitive, or
destructive interactions. Notice that $P_{ab}$ is assumed
to be independent of the position (i.e., translation invariant), while the
total payoff $\sum_{b}P_{ab}\,n_{i}^{b}(t)$ due to interactions 
depends on the distribution of entities over the system. 
The latter is an essential point for the possibility of 
fluctuation-induced self-organization. We also point out that,
in formula (\ref{formu}), pair interactions are restricted to 
the cell in which the
individual is located. Therefore, we do not assume spin-like or Ising-like
interactions, 
in contrast to other 
quantitative models proposed for the
description of social behavior \cite{Lewenstein,Galam}. 
\par
The quantities $\xi_\alpha(t)$ are random 
variables allowing to consider individual
variations of the success, which may be ``real'' or due to uncertainty
in the evaluation or estimation of success.
In our simulation program, they are uniformly 
distributed in the interval $[0,D_a]$, where $D_a$ is the
fluctuation strength (not to be mixed up wich a diffusion constant).
However, other specifications of the noise term are possible as well.
\par 
{\em 2nd step:\/} Determine the (expected) successes $S_a(i\pm 1,t)$
for the nearest neighbors $(i\pm 1)$ as well.
\par
{\em 3rd step:\/} Keep entity $\alpha$ in its previous cell $i$,
if $S_a(i,t) \ge \max \{S(i-1,t),S(i+1,t)\}$. Otherwise, move to cell
$(i-1)$, if $S(i-1,t) > S(i+1,t)$, and move to cell $(i+1)$, 
if $S(i-1,t) < S(i+1,t)$. In the remaining case $S(i-1,t) = S(i+1,t)$,
jump randomly to cell $(i-1)$ or $(i+1)$ with probability 1/2.
\par
If there is a maximum density $\rho_{\rm max} = N_{\rm max}/I$
of entities, overcrowding can be avoided by introducing a
saturation factor
\begin{equation}
 c(j,t) = 1 - \frac{N_j(t)}{N_{\rm max}}\, , \quad
 N_j(t) = \sum_a n_j^a(t) \, ,
\end{equation}
and performing the update steps with the generalized success 
\begin{equation}
 S'_a(j,t) = c(j,t) S_a(j,t)
\end{equation}
instead of $S_a(j,t)$, where $j \in \{i-1,i,i+1\}$. The model can be also
easily extended to include long distance interactions, 
jumps to more remote cells, etc. (cf. Section 5). 

\section{\large Simulation Results}

We consider two subpopulations $a\in\{1,2\}$ and $N_1 = N_2 = 100$
entities in each subpopulation, which are distributed over $I=20$ cells.
The payoff matrix $(P_{ab})$ will be represented by the vector 
$\vec{P} = (P_{11}, P_{12}, P_{21}, P_{22})$, 
where we will restrict ourselves
to $|P_{ab}|\in\{ 1, 2\}$ for didactical reasons.
For symmetric interactions between subpopulations, we have
$P_{ab} = P_{ba}$, while for 
asymmetric interactions, there is $P_{ab} \ne P_{ba}$, if $a\ne b$. 
For brevity, the interactions within the same population will be called
self-interactions, those between different populations cross-interactions. 
\par
To characterize the level of self-organization in each subpopulation
$a$, we can, for example, use the 
overall successes 
\begin{equation}
S_a(t) =  \frac{1}{I^2} \sum_i \sum_b n_i^a(t) \, P_{ab} \, n_i^b(t) \, , 
\end{equation}
the variances 
\begin{equation}
V_a(t) = \frac{1} {I^2} \sum_i \, [n^a_i(t)-n^a_{\rm hom}]^2 \, ,
\end{equation}
or the alternation strengths 
\begin{equation}
A_a(t) = \frac{1} {I^2} \sum_i \, [n^a_i(t)-n^a_{i-1}(t)]^2 \, .
\end{equation}

\subsection*{\normalsize 3.1. Symmetric Interactions}

By analogy with a more complicated model \cite{helvic} it is expected 
that the global overall success $S(t)=\sum_a S_a(t)$
is an increasing function in time, if the
fluctuation strengths $D_a$ are zero. However, what happens at finite noise
amplitudes $D_a$ is not exactly known. One would usually expect that finite
noise tends to obstruct or suppress self-organization, which will
be investigated in the following. 
\par
We start with the payoff matrix $\vec{P}=(2,-1,-1,2)$ corresponding to 
positive (or attractive) self-interactions and negative 
(or repulsive) cross-interactions. That is, entities of the same
subpopulation like each other, while entities of different
subpopulations dislike each other. The result will naturally be
segregation (``ghetto formation'') \cite{Sche71,We91}, 
if the noise amplitude is small.
However, segregation is suppressed by large fluctuations,
as expected (see Fig.~\ref{fig1}).
\par
\abb{\hspace*{-2mm}\includegraphics[height=5.5cm, angle=-90]{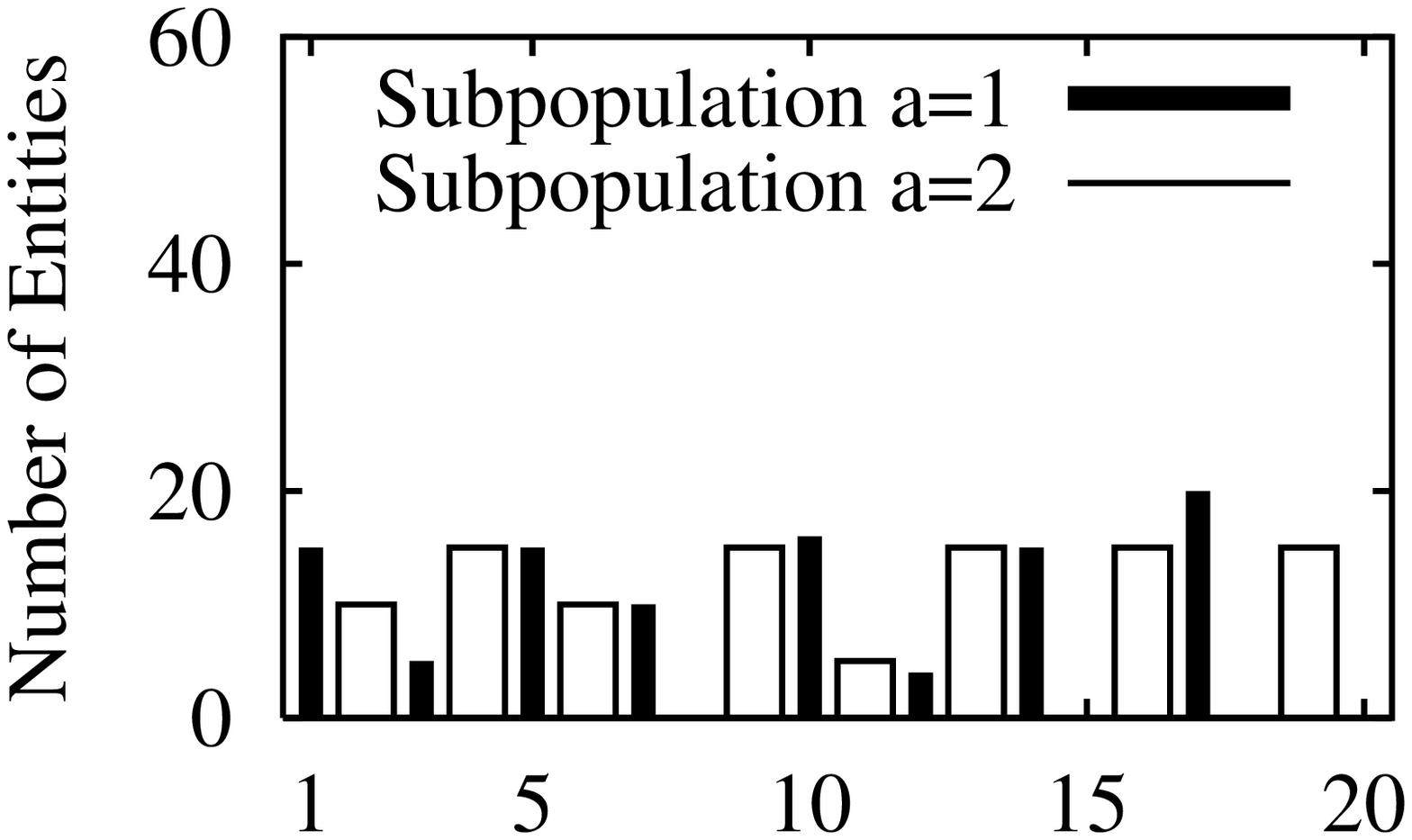}\\[-7.5mm]
\hspace*{-2mm}\includegraphics[height=5.5cm, angle=-90]{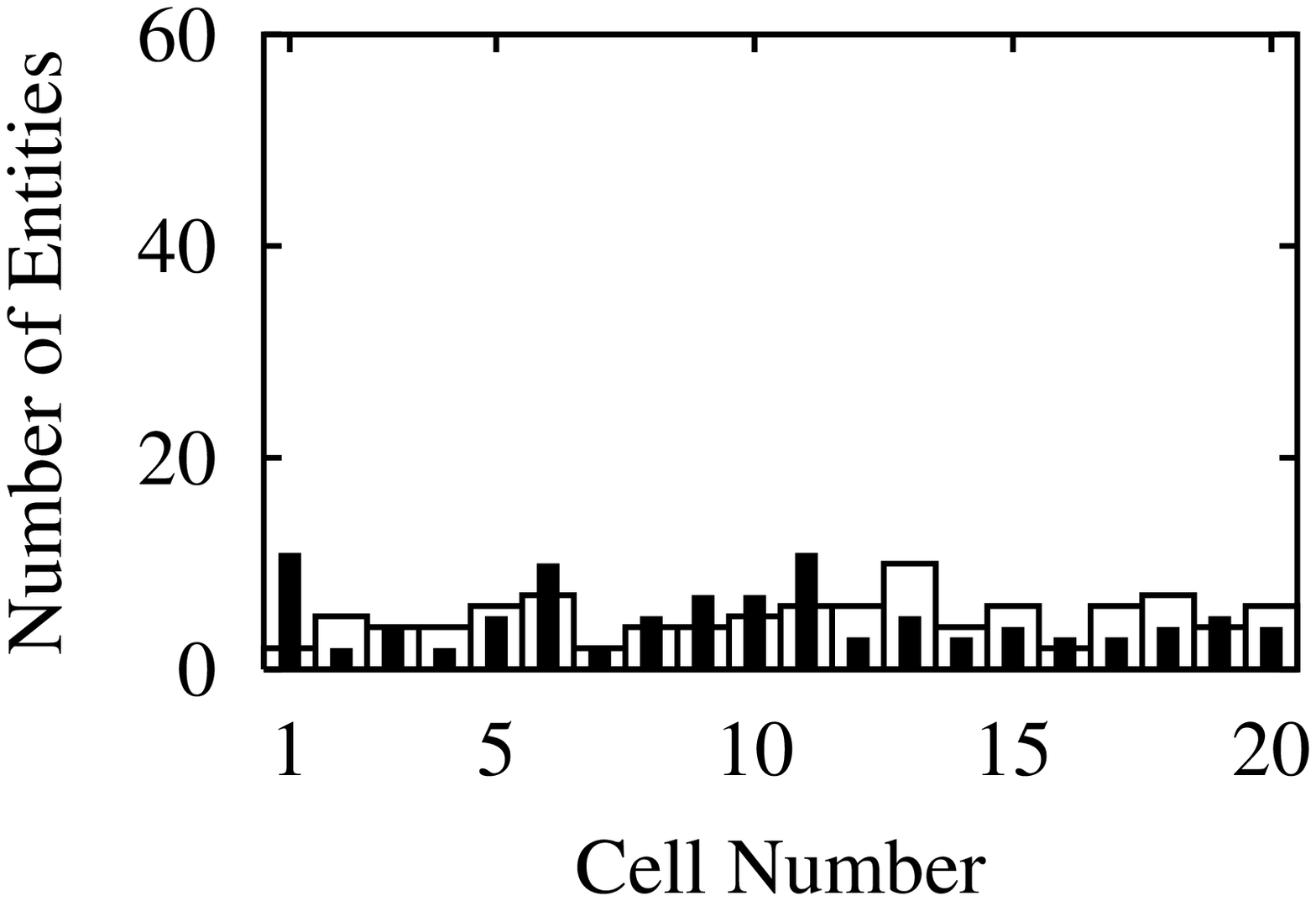}}
{Resulting distribution of entities at $t=4000$ for
the payoff matrix $\vec{P}=(2,-1,-1,2)$ at small fluctuation strength
$D_a=0.1$ (top) and large fluctuations strength $D_a=5$ (bottom).}
{fig1}
However, for medium noise amplitudes $D_a$, we find a much more pronounced
self-organization (segregation) than for small ones (compare
Fig.~\ref{fig2} with Fig.~\ref{fig1}). The effect is systematic
insofar as the degree of segregation (and, hence, the overall success)
increases with increasing noise amplitude, until segregation breaks
down above a certain critical noise level. 
\par\abb{\includegraphics[height=6.0cm, angle=-90]{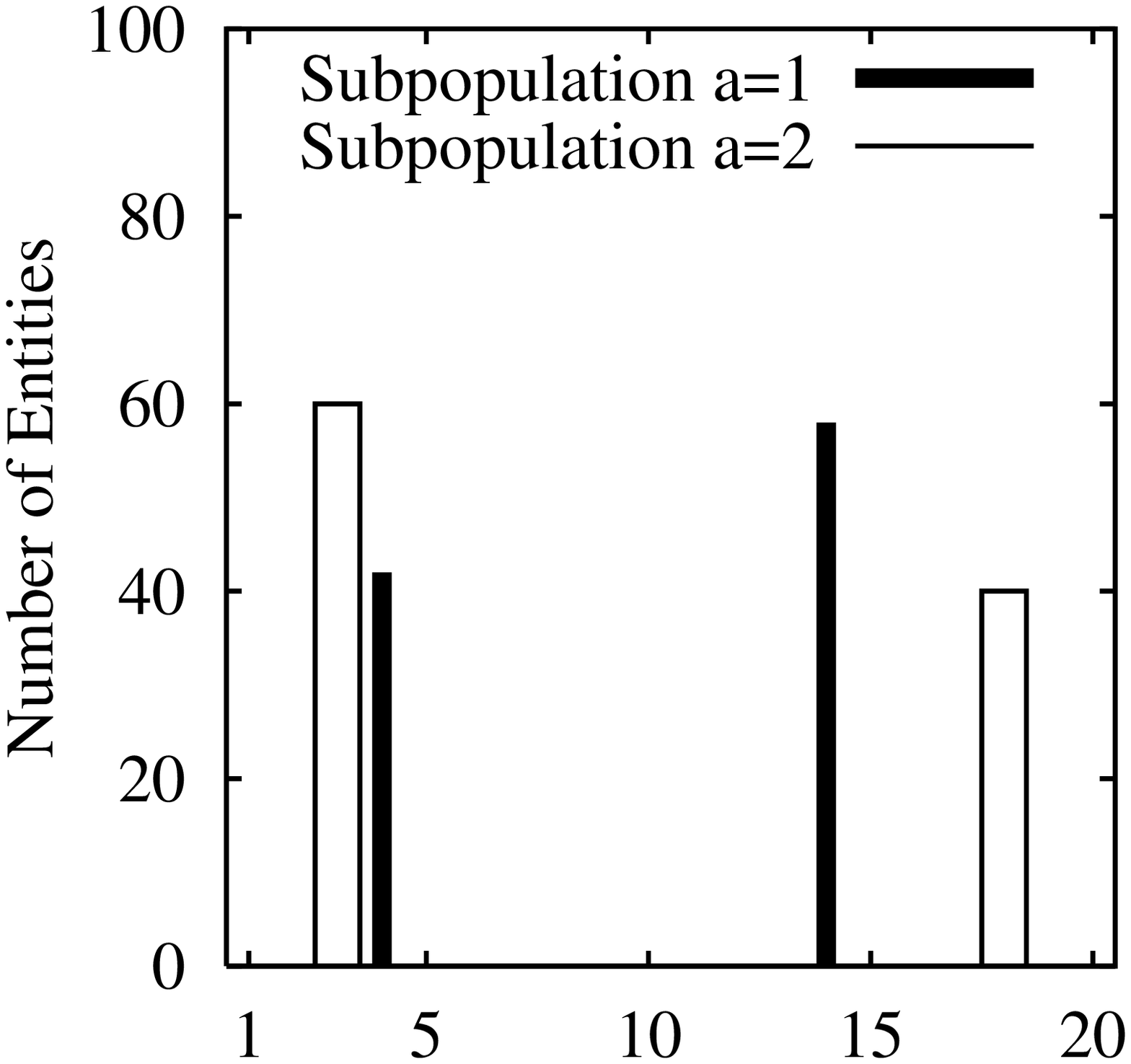}\\[-7.5mm]
\includegraphics[height=6.0cm, angle=-90]{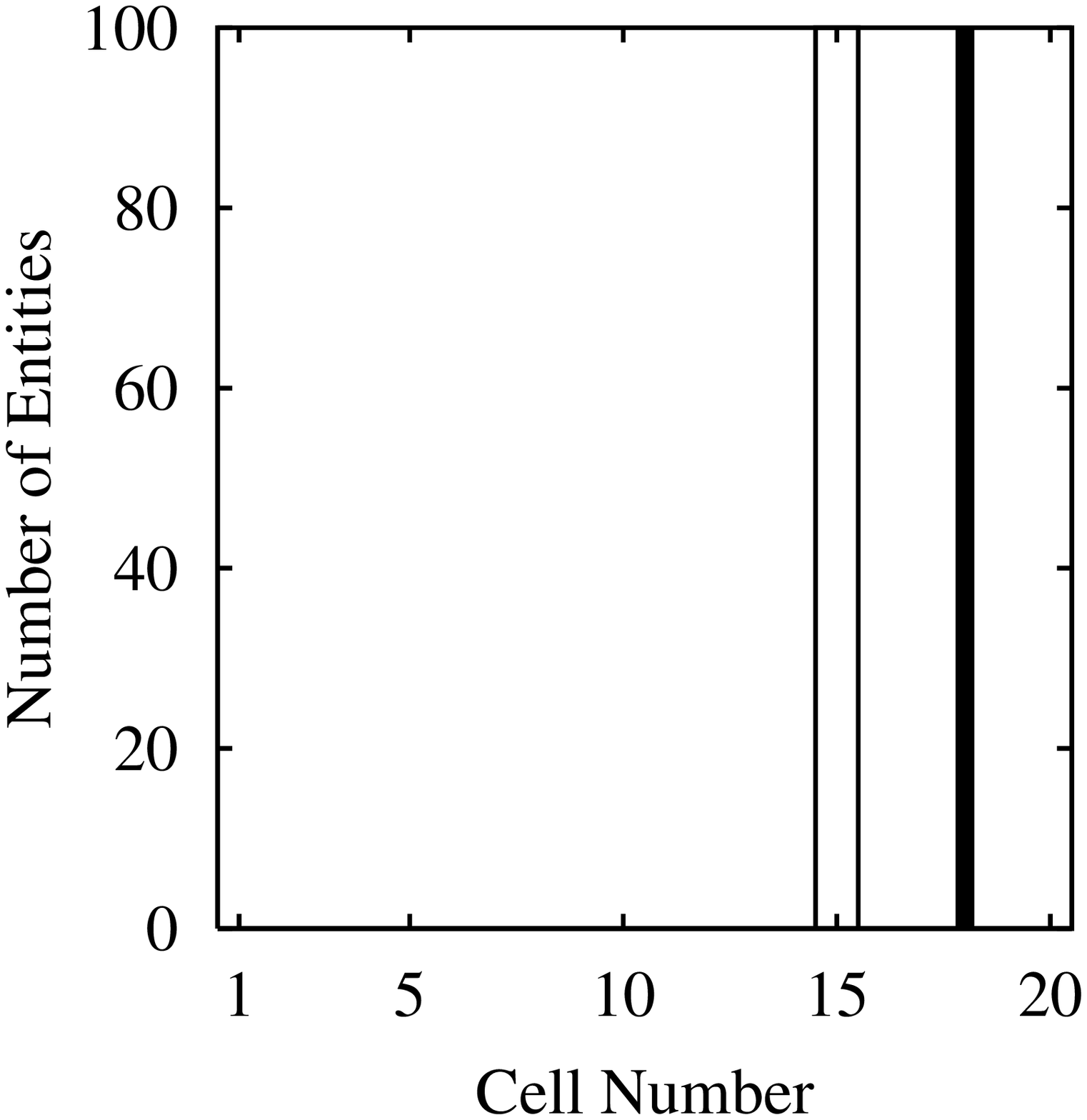}}
{As Fig.~\ref{fig1}, but with medium fluctuation strength
$D_a=2$ (top) and $D_a=3$ (bottom).}
{fig2}
Let us investigate some other cases: For the structurally similar 
payoff matrix $(1,-2,-2,1)$,
we find segregation as well, which is not surprising. In contrast, we
find agglomeration for the payoff matrices
$(1,2,2,1)$ and $(2,1,1,2)$. This agrees with intuition, since all
entities like each other in these cases, which makes them move to
the same places, like in the formation of settlements \cite{We91}, the
development of trail systems \cite{trail,biosys,millonas}, or the build up of
slime molds \cite{biol6,Pri2}. More interesting is the case corresponding to 
the payoff matrix $(-1,2,2,-1)$, where the cross-interactions are positive
(attractive), while the self-interactions are
negative (repulsive). One may think that this causes the entities of the same
subpopulation to spread homogeneously over the system, 
and in all cells would result an equal number of entities 
of both subpopulations, which is compatible with mutual attraction.
However, this homogeneous distribution turns out to be unstable with
respect to fluctuations. Instead, we find agglomeration! This result
is more intuitive if we imagine one subpopulation to represent
women and the other one men (without taking this example too serious). 
While the interaction between women
and men is normally strongly attractive, the interactions among men or
among women may be considered to be weakly competitive. As we all
know, the result is a tendency of young men and women to move into 
cities. Corresponding simulation results for different noise strengths
are depicted in Fig.~\ref{fig3}. Again, we find that the
self-organized pattern is destroyed by strong fluctuations
in favour of a more or less 
homogeneous distribution, while medium noise strengths
further self-organization. 
\par\abb{
\hspace*{-2mm}\includegraphics[height=5.5cm, angle=-90]{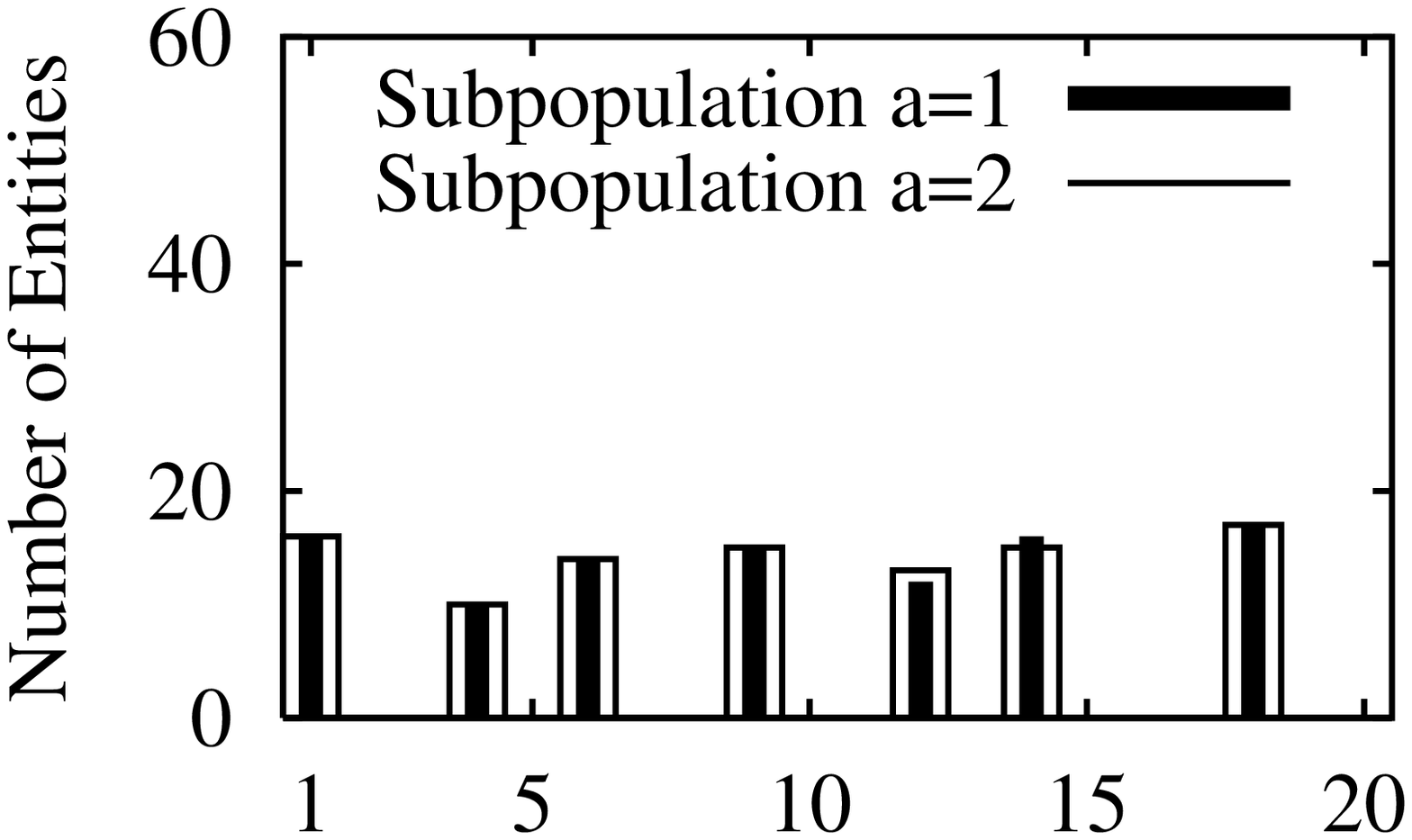}\\[-7.5mm]
\includegraphics[height=6.0cm, angle=-90]{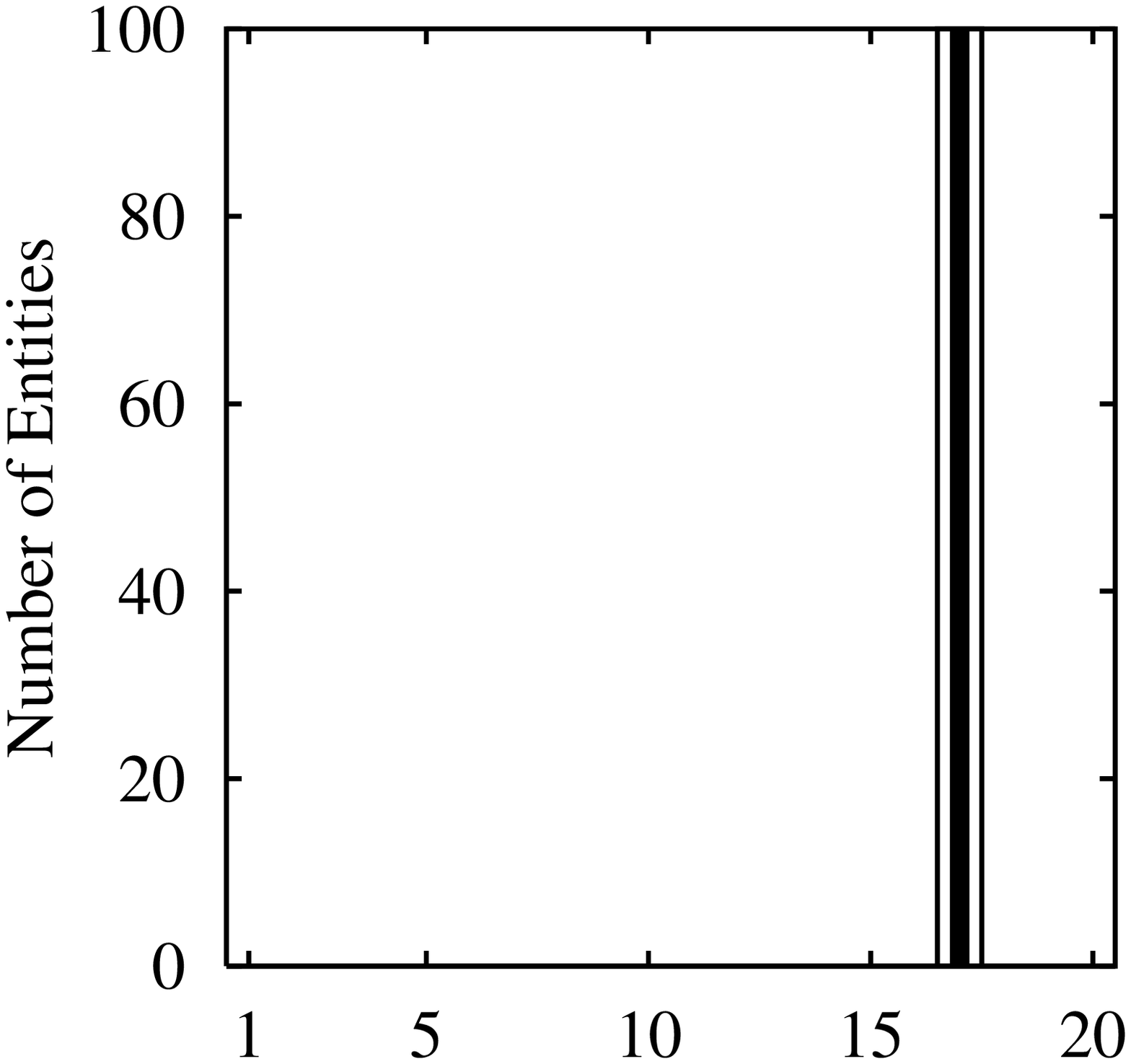}\\[-7.5mm]
\hspace*{-2mm}\includegraphics[height=5.5cm, angle=-90]{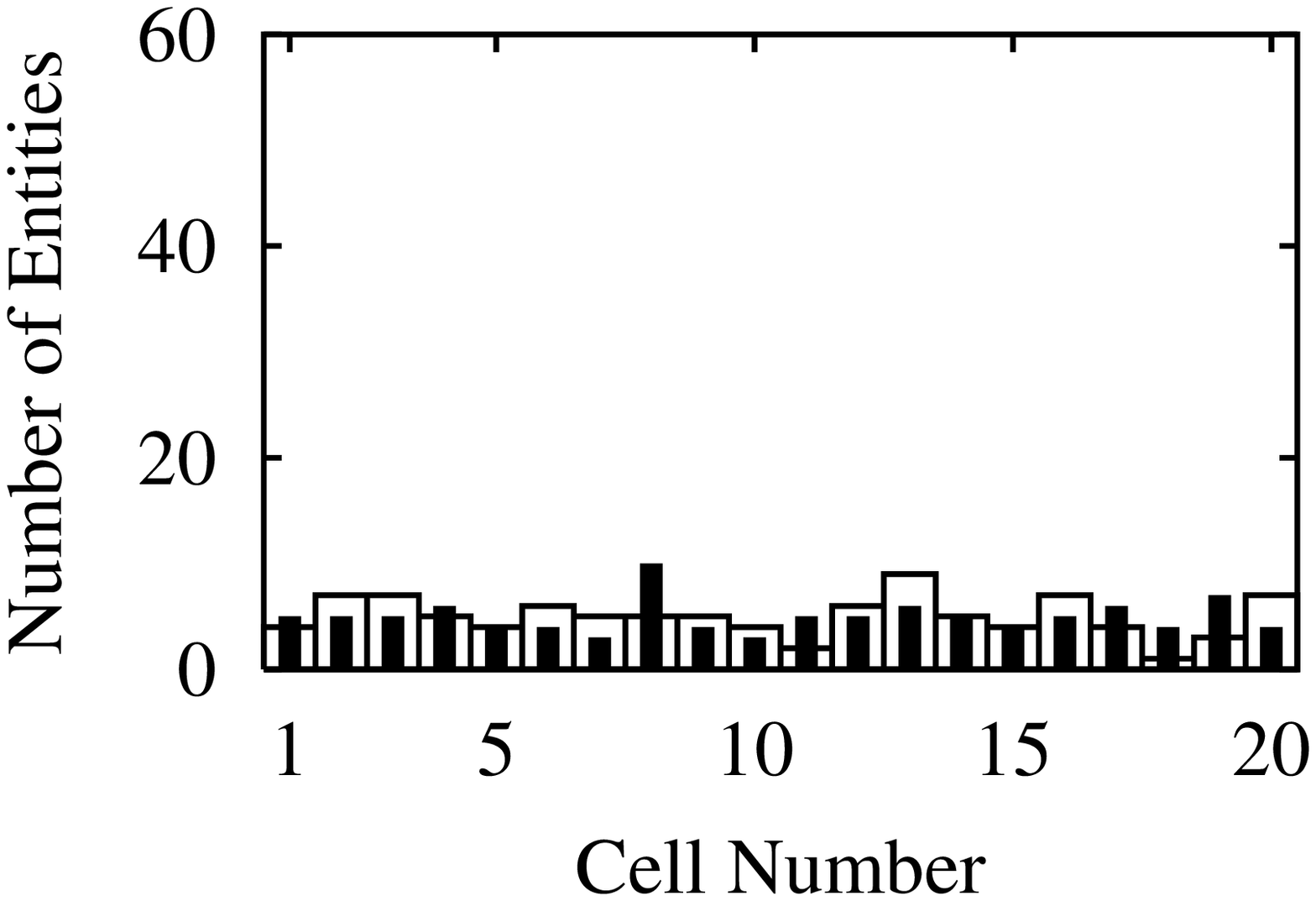}}
{As Fig.~\ref{fig1}, but for the payoff matrix 
$\vec{P}=(-1,2,2,-1)$ and 
$D_a=0.05$ (top), $D_a=1.5$ (middle), and $D_a=5$ (bottom).}
{fig3}
For the payoff matrices $(-2,1,1,-2)$ and $(-2,-1,-1,-2)$, i.e. cases
of strong negative self-interactions, we find
a more or less homogeneous distribution of entities in both
subpopulations, irrespective of the noise amplitude. In contrast,
the payoff matrix $(-1,-2,-2,-1)$ corresponding to negative
self-interactions but even stronger negative cross-interactions,
leads to another self-organized pattern. We may describe it
as the formation of lanes, as it is observed in pedestrian counterflows
\cite{pre,helvic} or in sheared granular media with different
kinds of grains \cite{granular}. 
While both subpopulations tend to separate from
each other, at the same time they tend to spread over all the
available space (see Fig.~\ref{fig4}), 
in contrast to the situation depicted in Figs.~\ref{fig1} and
\ref{fig2}. Astonishingly enough, a medium level of noise again
supports self-organized ordering, since it 
helps the subpopulations to separate from each other.
\par\abb{
\hspace*{-2mm}\includegraphics[height=5.5cm, angle=-90]{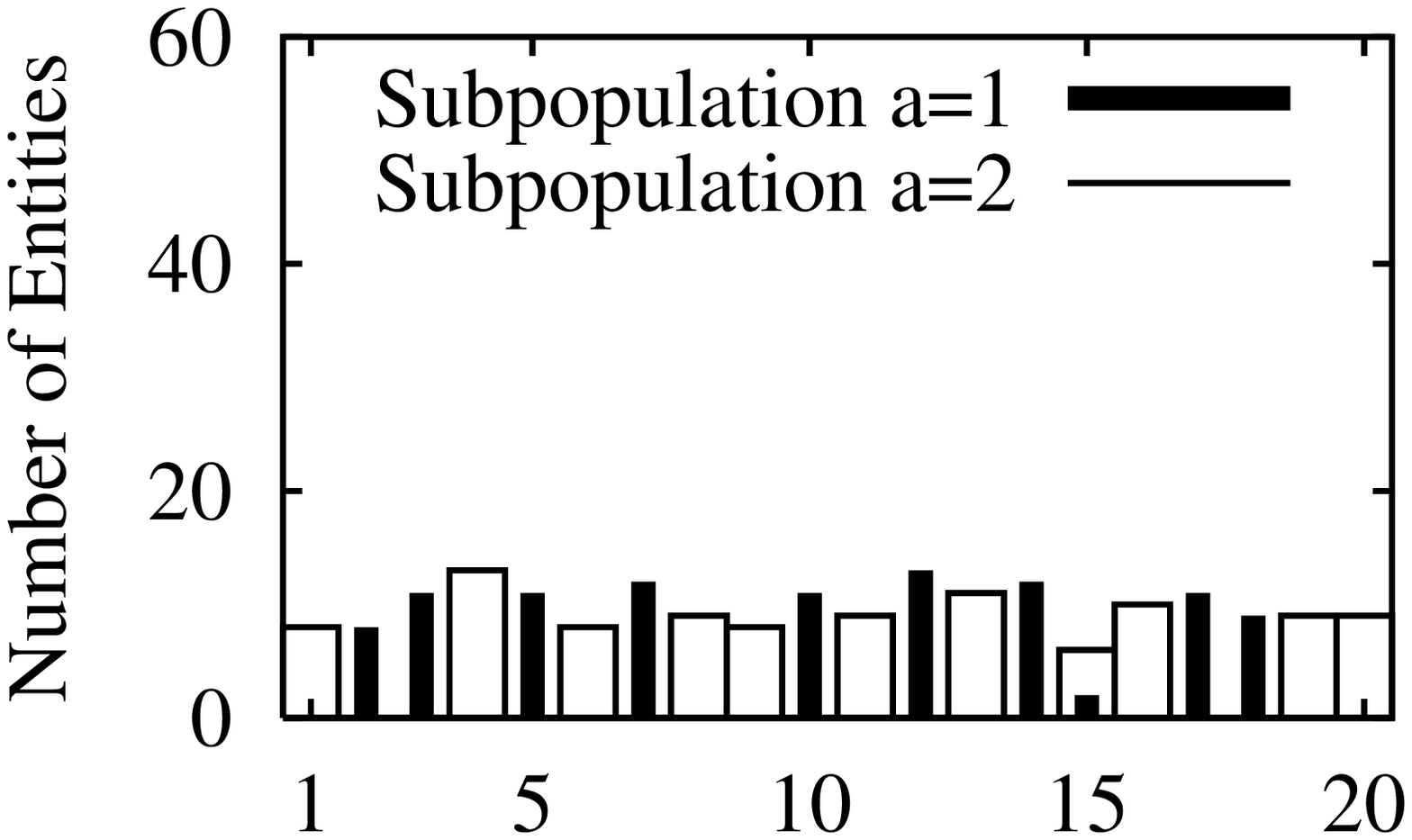}\\[-7.5mm]
\hspace*{-2mm}\includegraphics[height=5.5cm, angle=-90]{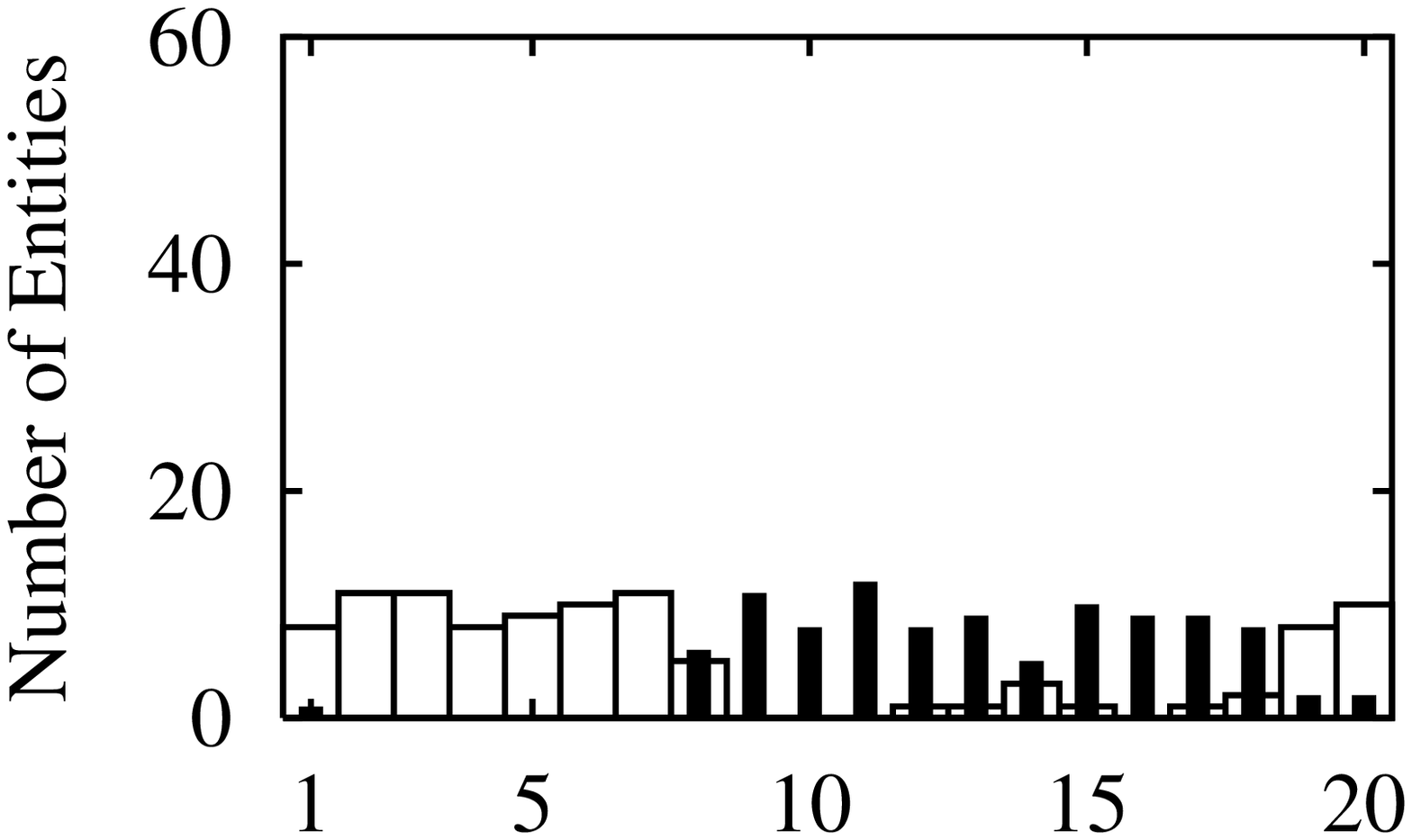}\\[-7.5mm]
\hspace*{-2mm}\includegraphics[height=5.5cm, angle=-90]{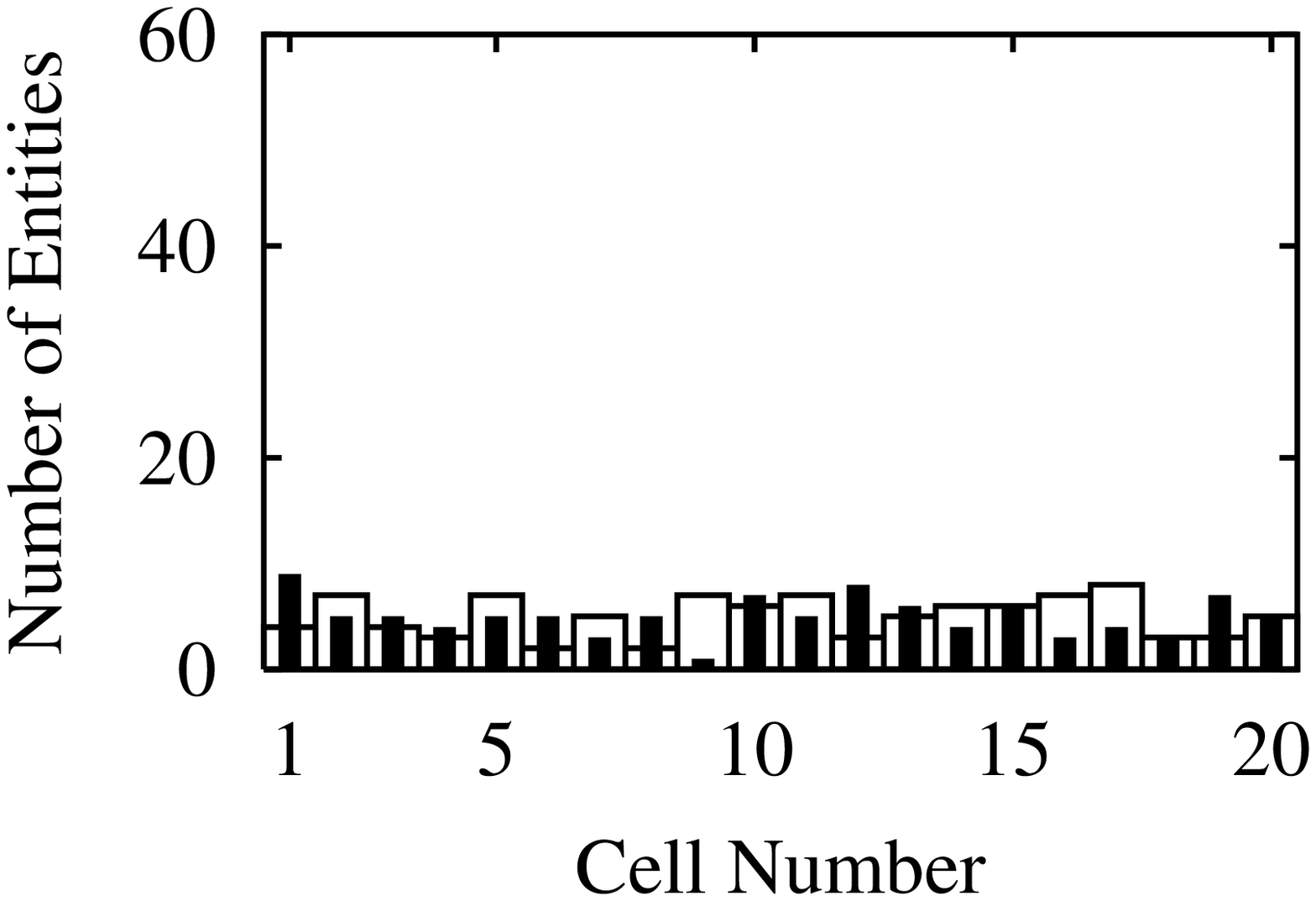}}
{As Fig.~\ref{fig1}, but for the payoff matrix 
$\vec{P}=(-1,-2,-2,-1)$ and 
$D_a=0.05$ (top), $D_a=0.5$ (middle), and $D_a=5$ (bottom).}
{fig4}
We finally mention  that a 
finite saturation level suppresses self-organization in a
surprisingly strong way, as is shown in Fig.~\ref{fig5}. Instead of
pronounced segregation, we will find a result similar to lane
formation, and even strong agglomeration will be replaced by 
an almost homogeneous distribution.
\par\abb{\hspace*{-2mm}\includegraphics[height=5.5cm, angle=-90]{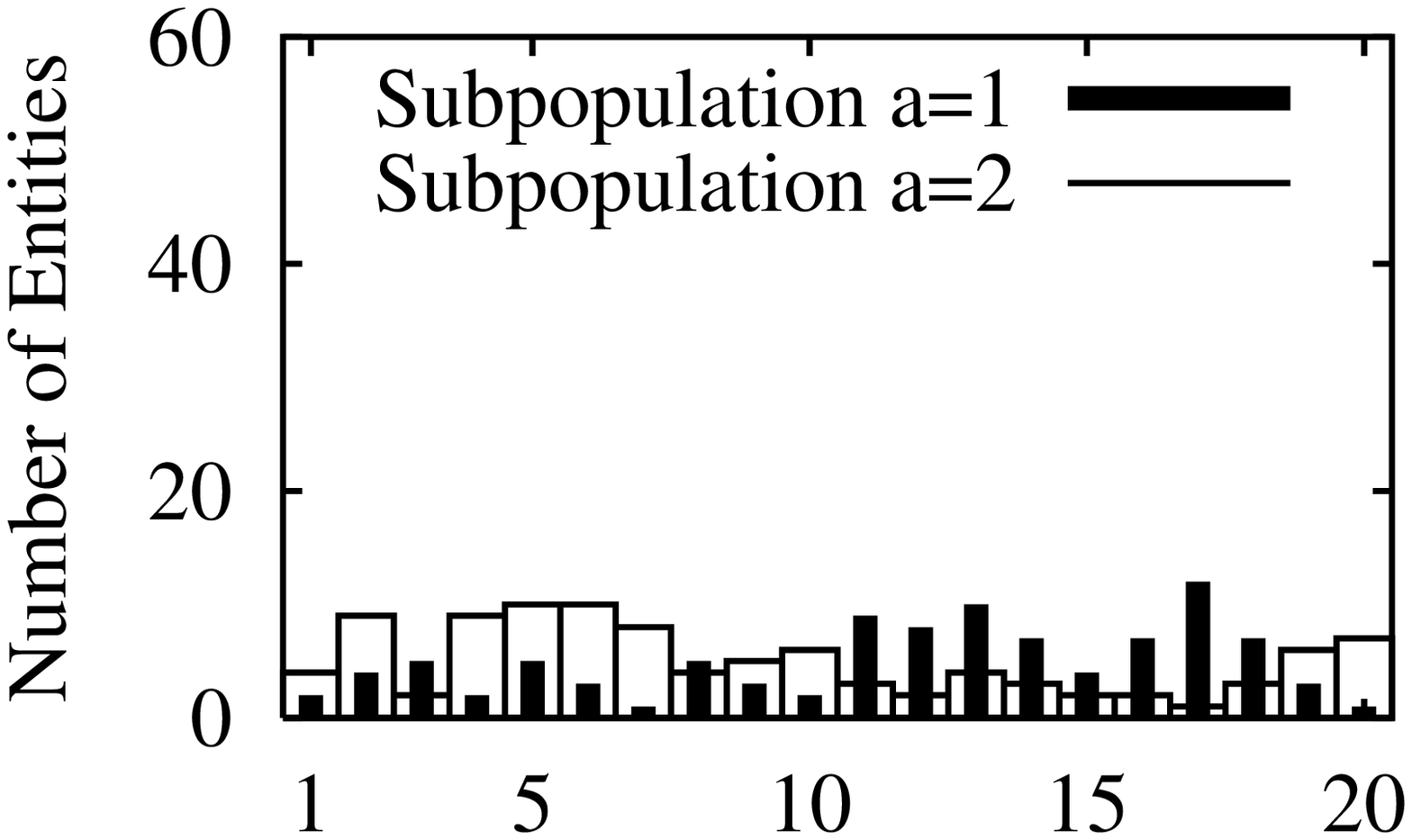}\\[-7.5mm]
\hspace*{-2mm}\includegraphics[height=5.5cm, angle=-90]{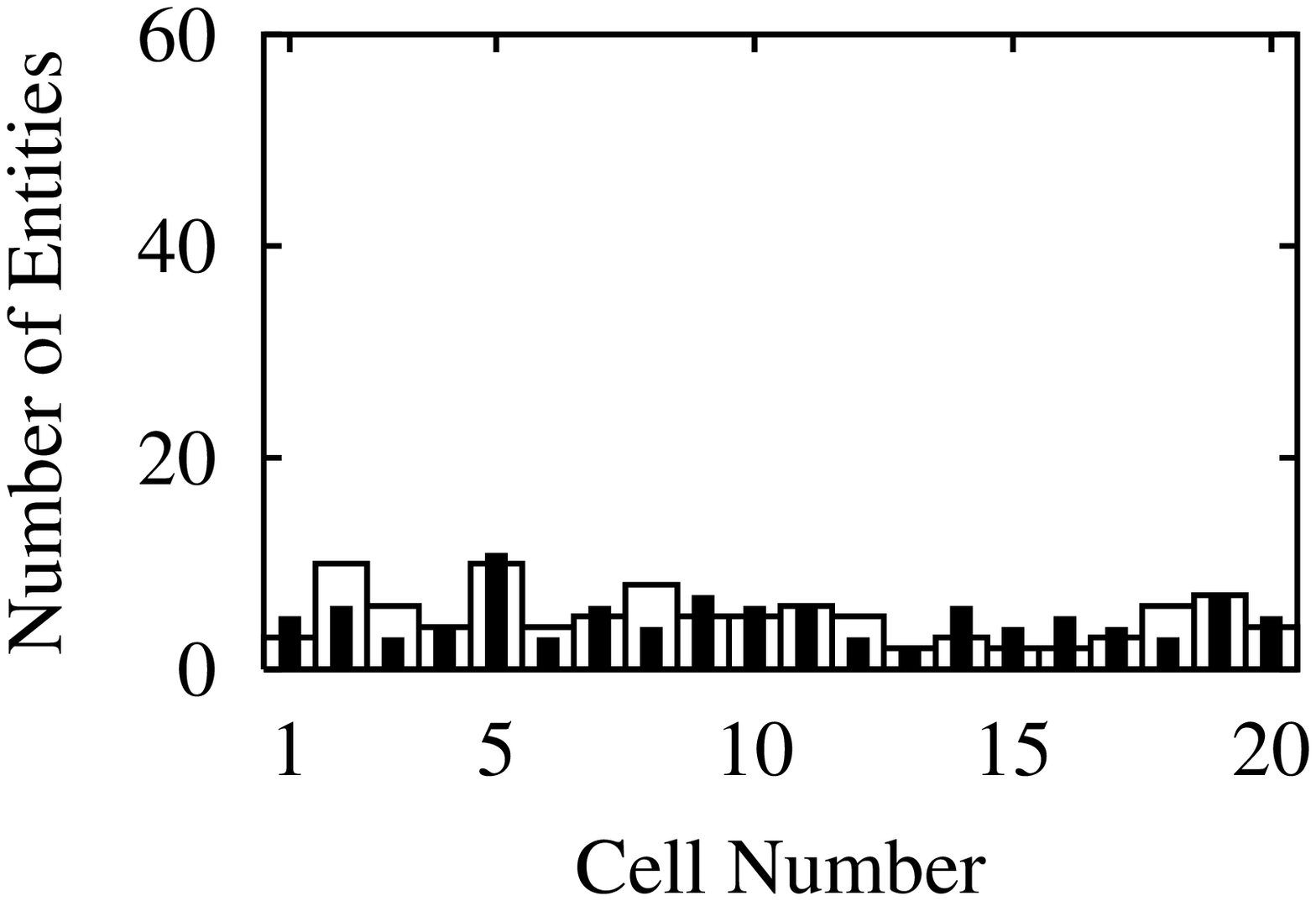}}
{Resulting distribution of entities at $t=4000$ with saturation level 
$N_{\rm max}=50$. Top: $\vec{P}=(2,-1,-1,2)$ and $D_a=3$.
Bottom: $\vec{P}=(-1,2,2,-1)$ and $D_a=1.5$.}
{fig5}

\subsubsection*{{\sss Noise-induced ordering}}
A possible interpretation for noise-induced ordering would be that 
fluctuations allow the system to leave local minima (corresponding to
partial agglomeration or segregation only). This could trigger
a transition to a more stable state with more pronounced
ordering. However, although this interpretation is consistent 
with a related example discussed in Ref.~\cite{helvic}, the idea of
a step-wise coarsening process is not supported by the
temporal evolution of the distribution of entities (see Fig.~\ref{fig6}) and
the time-dependence of the overall success within the subpopulations
(see Fig.~\ref{fig7}). This idea 
is anyway not applicable to segregation, since, 
in the one-dimensional case, the repulsive 
clusters of different subpopulations cannot simply pass each other in
order to join others of the same subpopulation.
\par
According to Figs.~\ref{fig6} and \ref{fig7},
segregation and agglomeration rather take place in
three phases: First, there is a certain time interval, during which
the distribution of entities remains more or less homogeneous. Second,
there is a short period of rapid self-organization. Third, there is
a continuing period, during which the distribution and overall success 
do not change anymore. The latter is a consequence of the short-range
interactions within our model, which are limited to the nearest
neighbors. Therefore, the segregation or aggregation process
practically stops, after separate peaks have evolved. This is not the
case for lane formation, where the entities redistribute, but
all cells remain occupied, so that we have ongoing interactions. This 
is reflected in the non-stationarity of the lanes and by the
oscillations of the overall success. 
\par\abb{
\vspace*{-5mm}
\hspace*{6mm}\includegraphics[height=6cm, angle=-90]{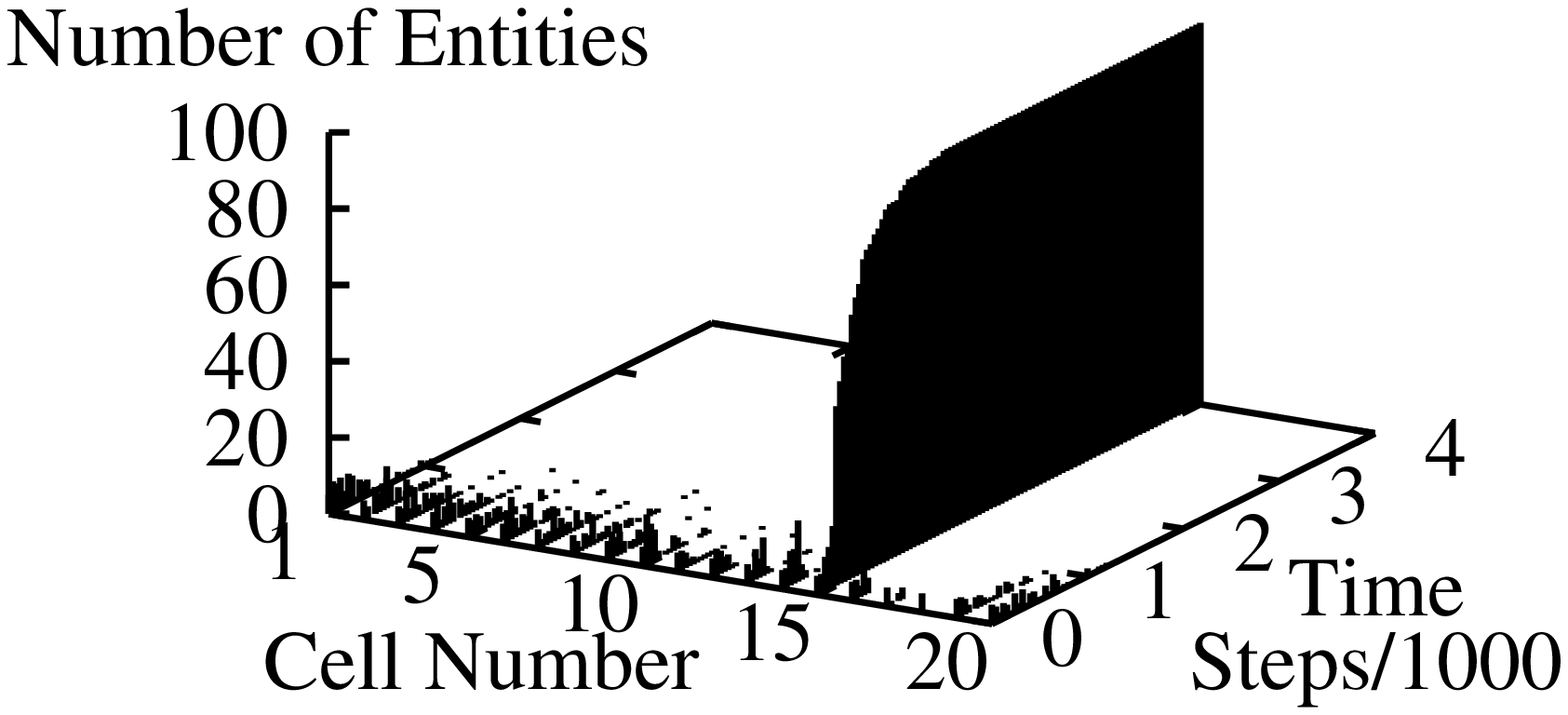}\\[-10mm]
\hspace*{6mm}\includegraphics[height=6cm, angle=-90]{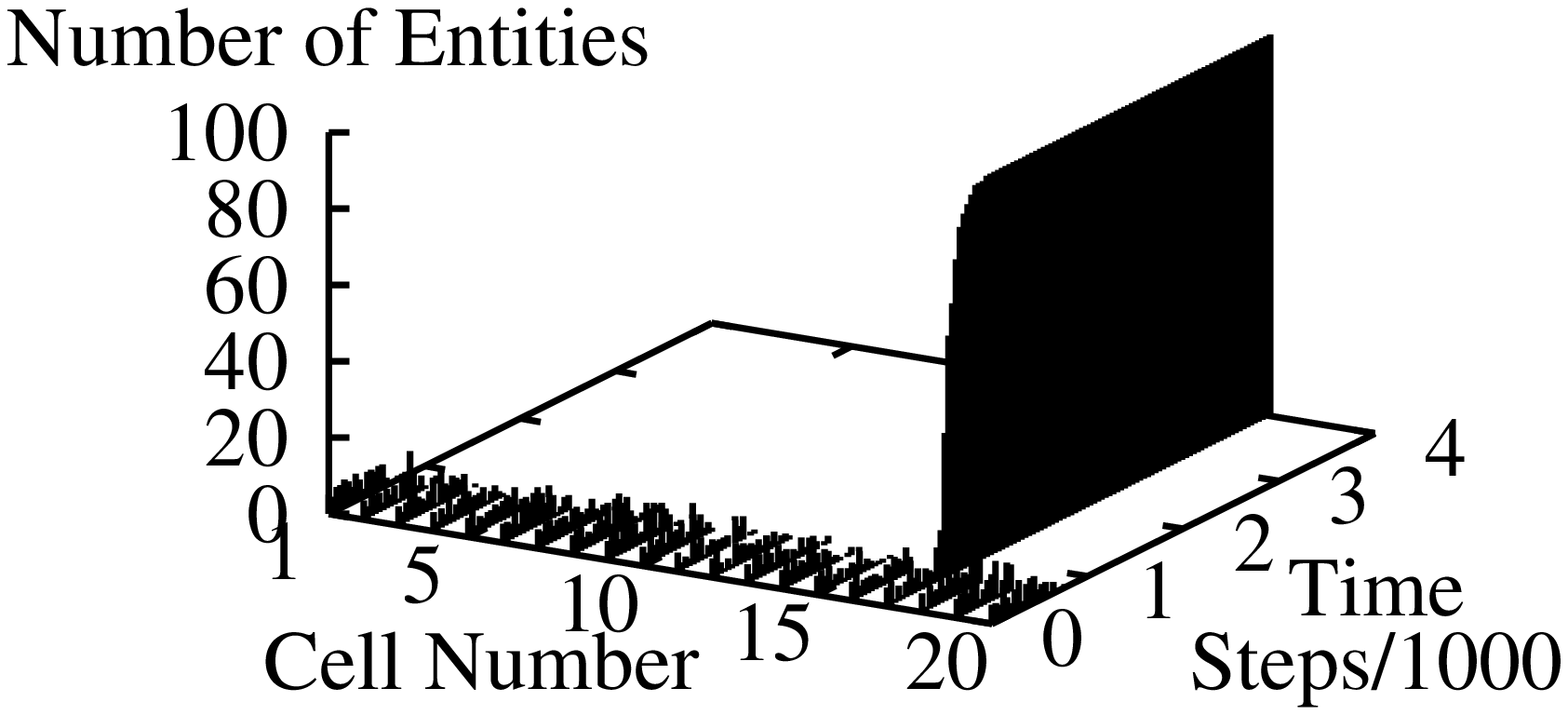}\\[-10mm]
\hspace*{6mm}\includegraphics[height=6cm, angle=-90]{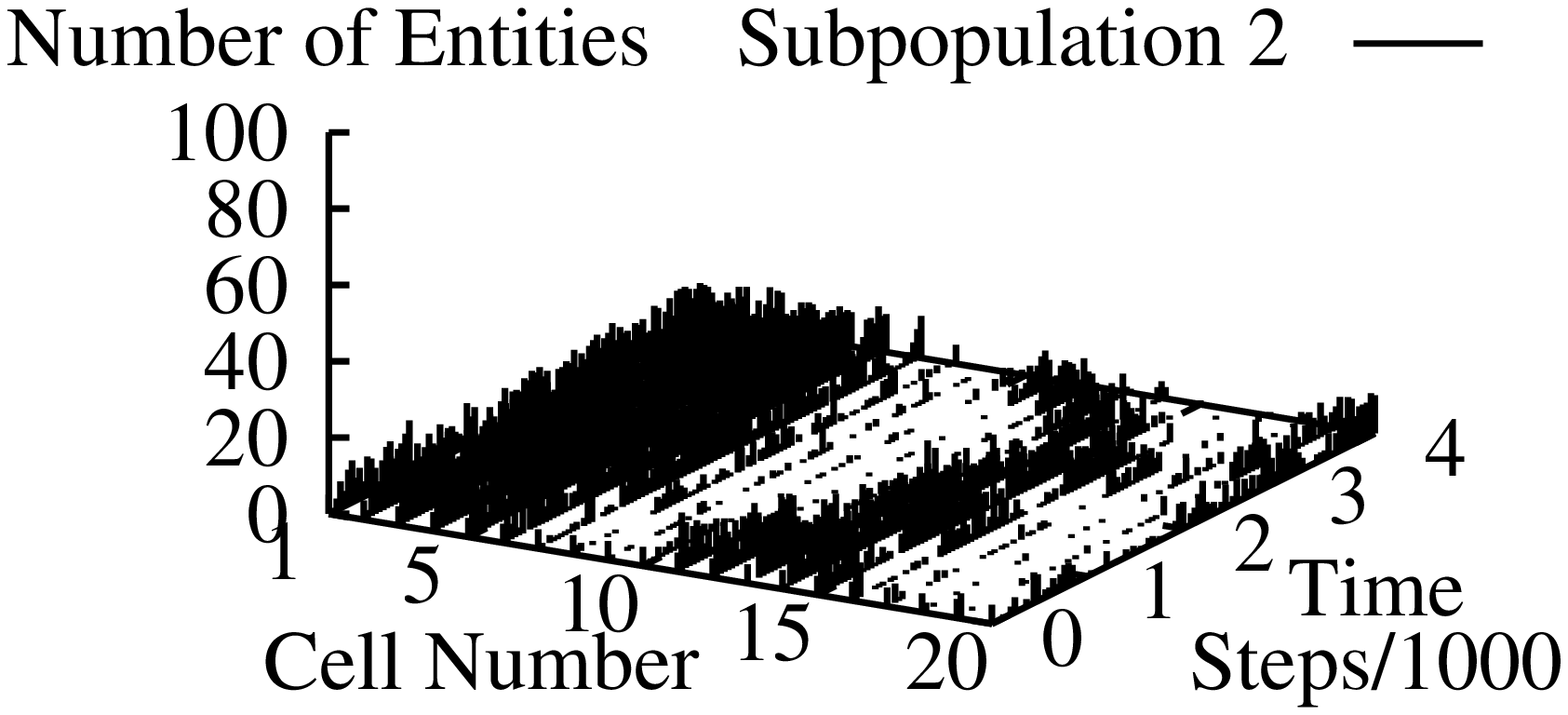}
\vspace*{-5mm}}
{Temporal evolution of the distribution of entitities within subpopulation
$a=2$ for $\vec{P}=(2,-1,-1,2)$ and $D_a=3$ (top), 
$\vec{P}=(-1,2,2,-1)$ and $D_a=1.5$ (middle), and
$\vec{P}=(-1,-2,-2,-1)$ and $D_a=0.5$ (bottom).}
{fig6}
\par\abb{
\hspace*{-2mm}\includegraphics[height=5.5cm, angle=-90]{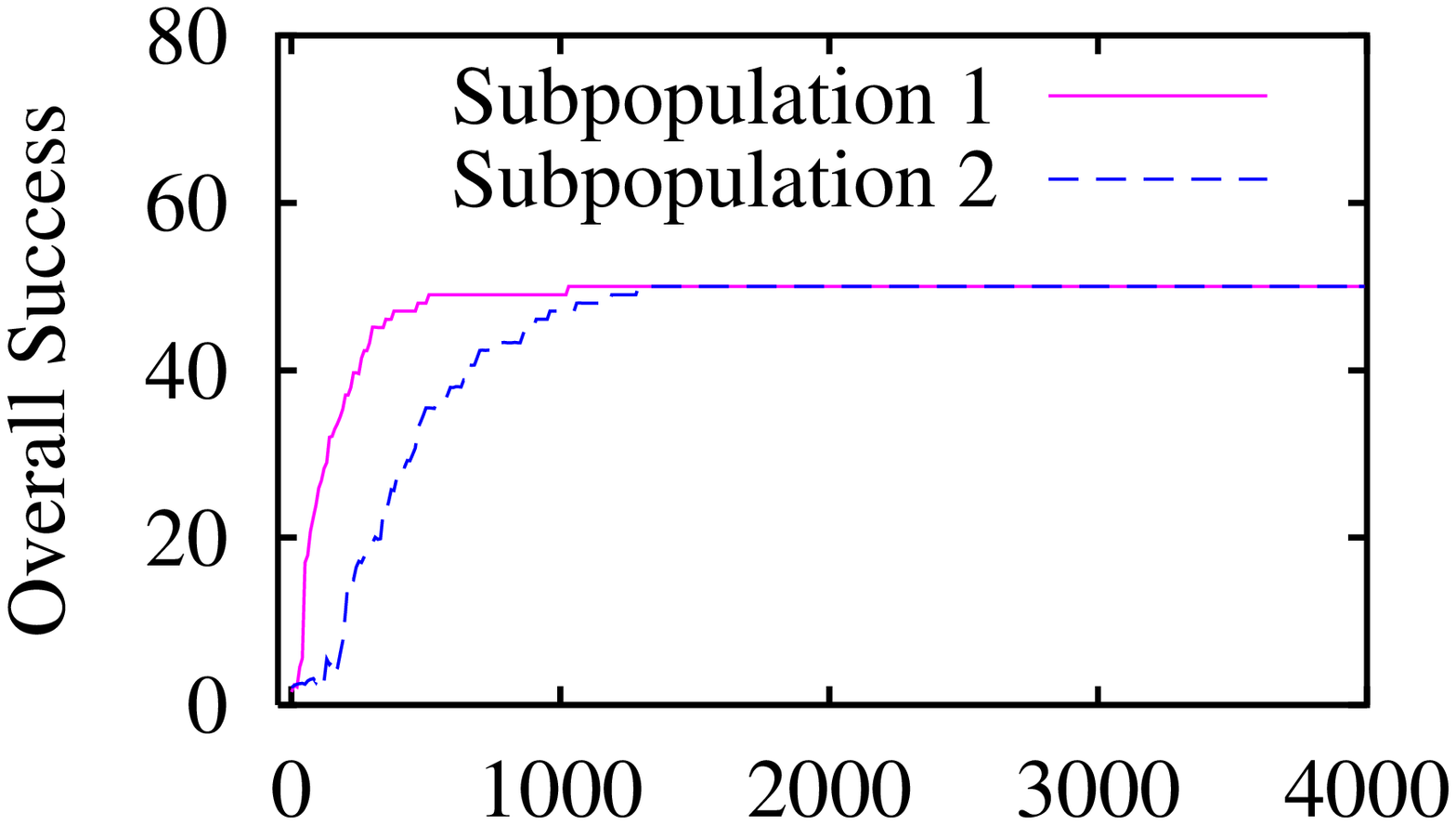}\\[-7.5mm]
\hspace*{-2mm}\includegraphics[height=5.5cm, angle=-90]{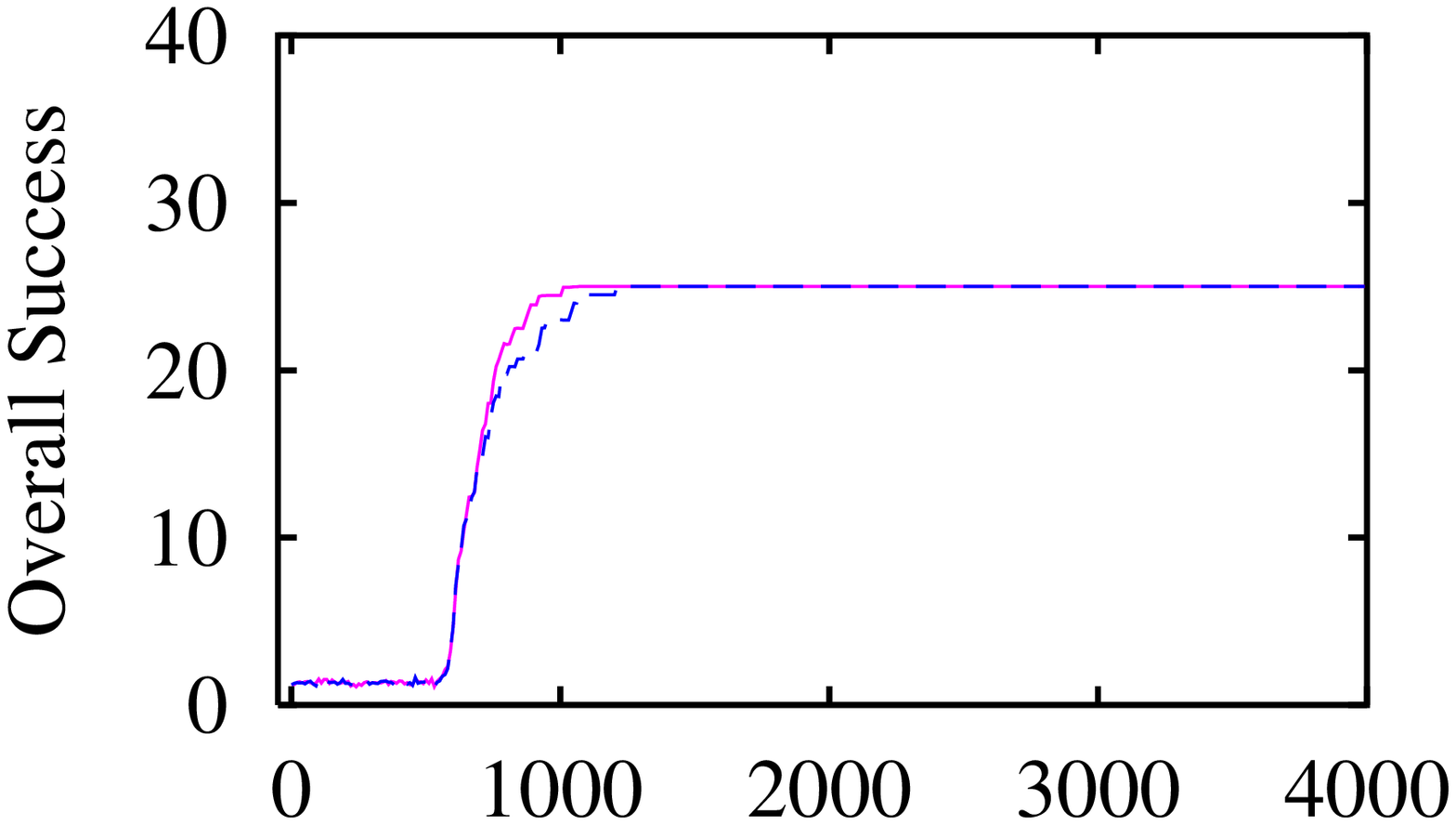}\\[-7.5mm]
\hspace*{-2mm}\includegraphics[height=5.55cm, angle=-90]{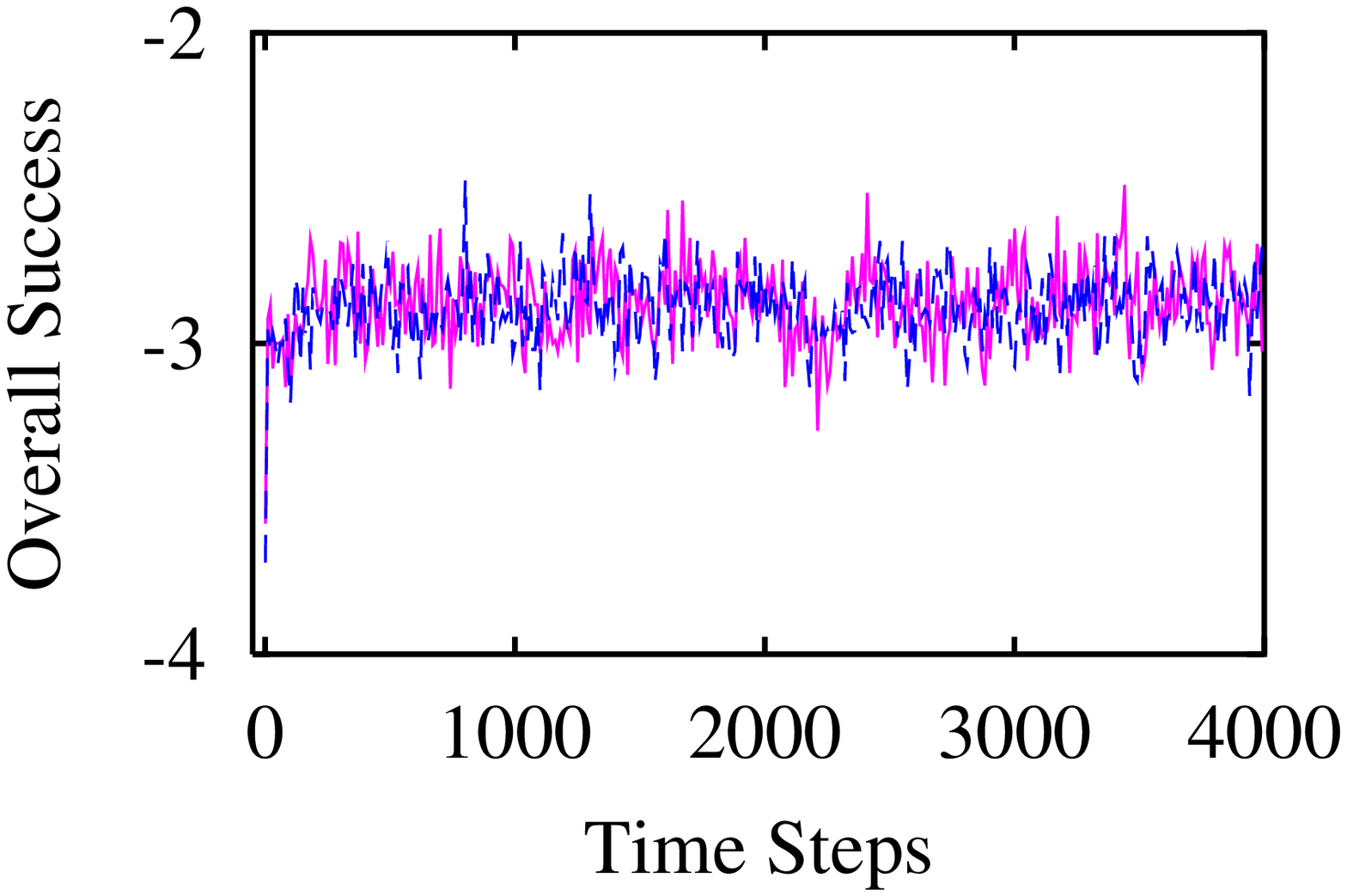}}
{Temporal evolution of the overall success within both subpopulations
for $\vec{P}=(2,-1,-1,2)$ and $D_a=3$ (top), 
$\vec{P}=(-1,2,2,-1)$ and $D_a=1.5$ (middle), and
$\vec{P}=(-1,-2,-2,-1)$ and $D_a=0.5$ (bottom).}
{fig7}
We suggest the following interpretation for the three phases mentioned 
above: 
During the first time interval, which is
characterized by a quasi-continous distribution of entities over
space, a long-range pre-ordering process takes place.
After this ``phase of preparation'',
order develops in the second phase similar to crystallization, and 
it persists in the third phase. The role of fluctuations seems to be the
following: An increased noise level avoids a rash local
self-organization by keeping up a quasi-continuous distribution of
entities, which is required for a redistribution of entities over
larger distances. In this way, a higher noise level increases the effective
interaction range by extending the first phase, the ``interaction phase''. 
As a consequence, the resulting structures are more extended in space
(but probably without a characteristic length scale, see
Introduction). 
\par
It would be interesting to investigate, whether this
mechanism has something to do with the recently discovered phenomenon
of ``freezing by heating'', where a medium noise level causes
a transition to a highly ordered (but energetically less stable)
state, while extreme noise levels
produce a disordered, homogeneous state again \cite{freezing}.

\subsection*{\normalsize 3.2. Asymmetric Interactions}

Even more intriguing transitions than in the symmetric case can be
found for 
asymmetric interactions between the subpopulations. Here,
we will focus on the payoff matrix $(-1,2,-2,1)$, only. This example
corresponds to the curious case, where individuals of subpopulation 1 
weakly dislike each other, but strongly like individuals of the
other subpopulation. In contrast, individuals of subpopulation 2 
weakly like each other, but they strongly dislike the other
subpopulation. A good example for this is hard to find. With some good 
will, one may imagine subpopulation 1 to represent poor
people, while subpopulation 2 corresponds to rich people. What will be 
the outcome? In simple terms, the rich are expected to agglomerate in
a few areas, if the poor are moving too nervously (see
Fig.~\ref{fig8}). In detail,
however, the situation is quite complex, as discussed in the next
paragraph.
\abb{
\hspace*{-2mm}\includegraphics[height=5.5cm, angle=-90]{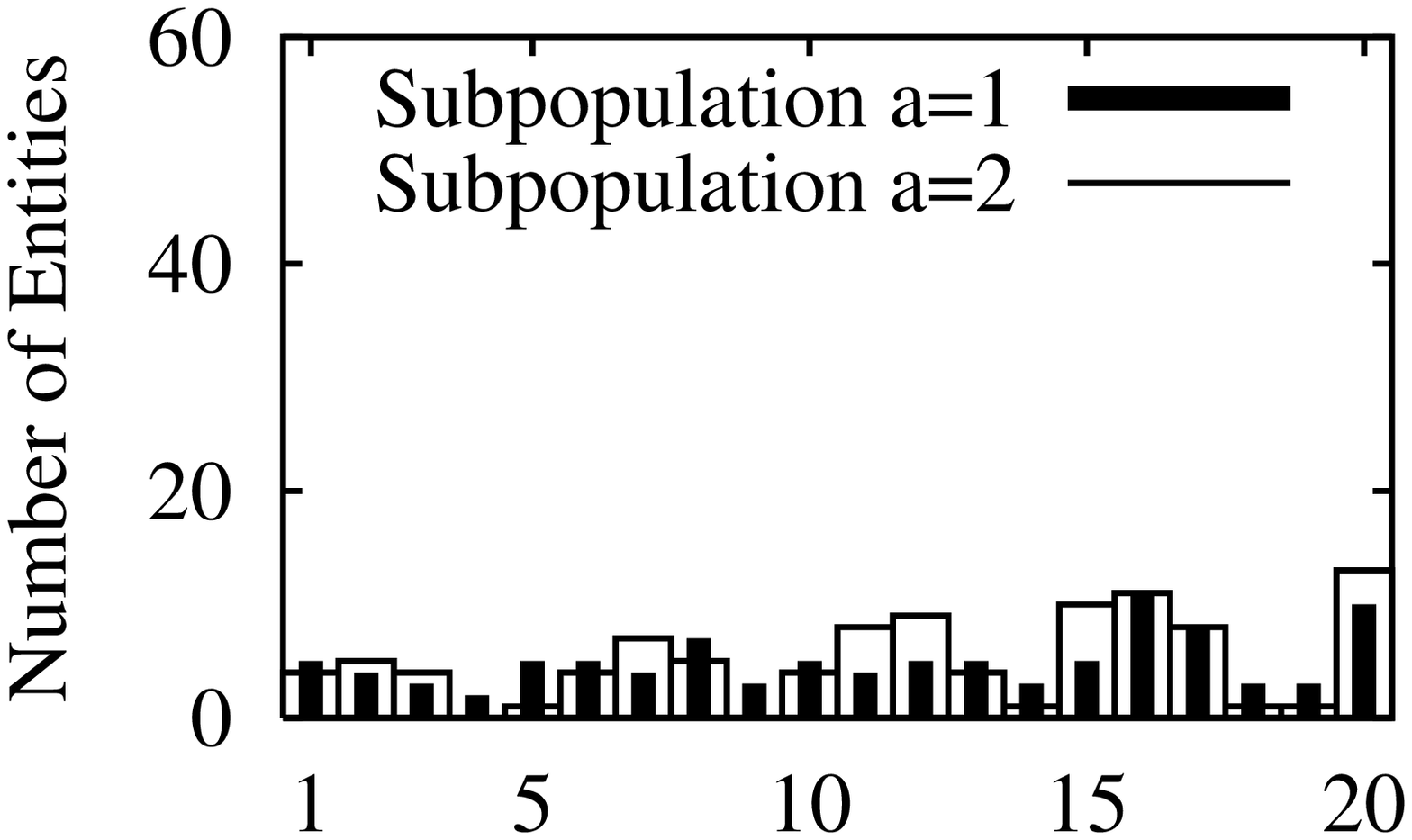}\\[-7.5mm]
\hspace*{-2mm}\includegraphics[height=5.5cm, angle=-90]{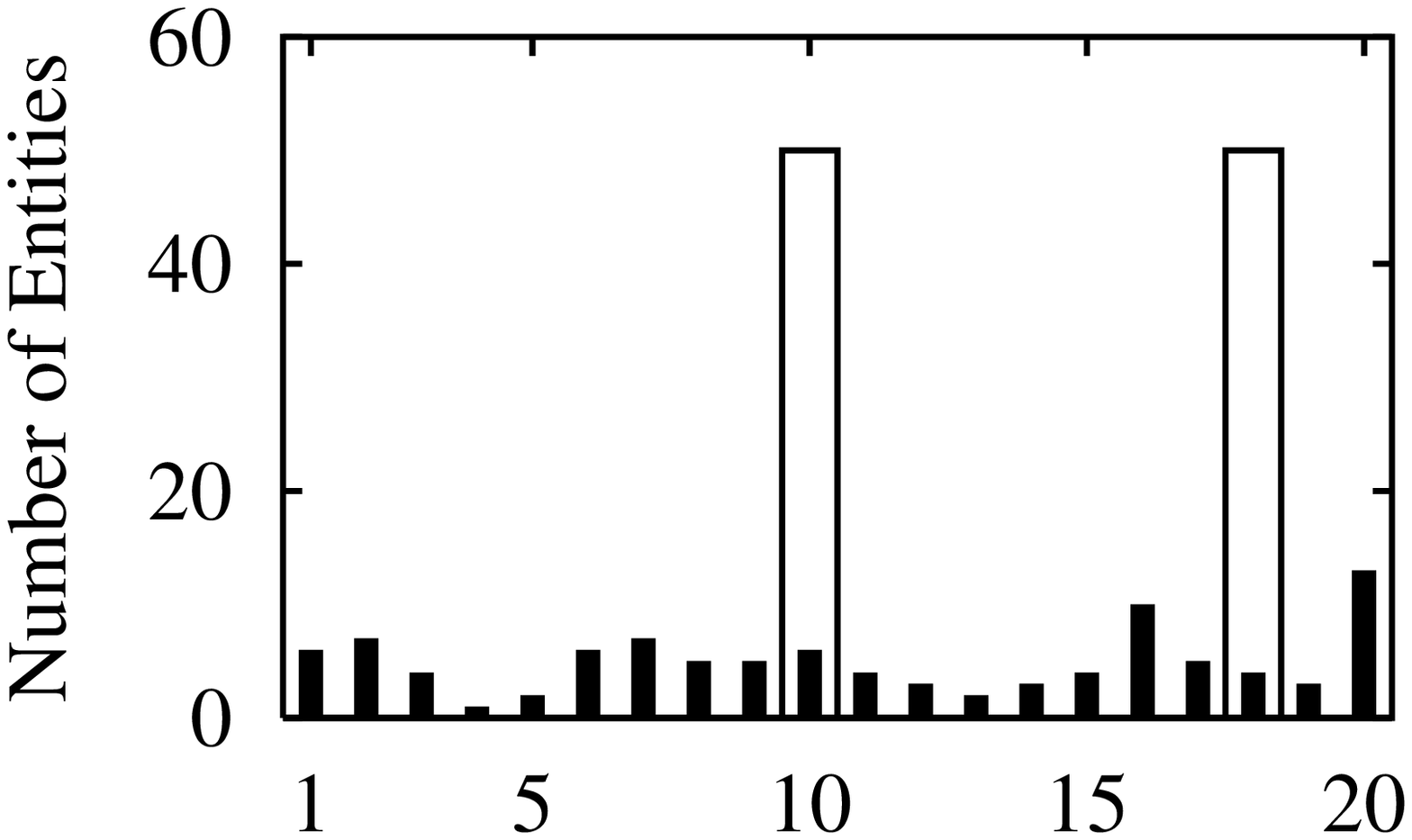}\\[-7.5mm]
\includegraphics[height=6.0cm, angle=-90]{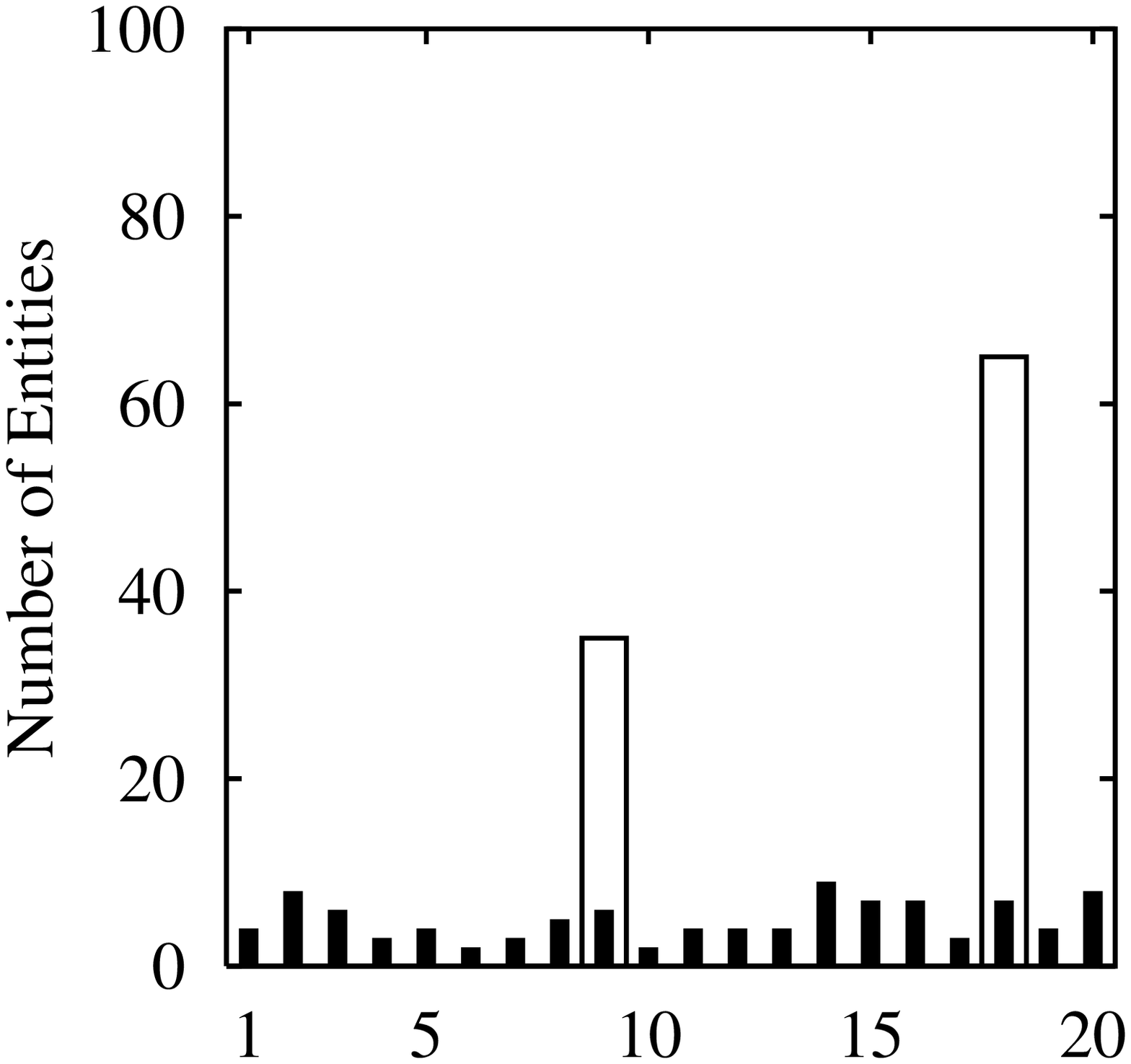}\\[-7.5mm]
\hspace*{-2mm}\includegraphics[height=5.5cm, angle=-90]{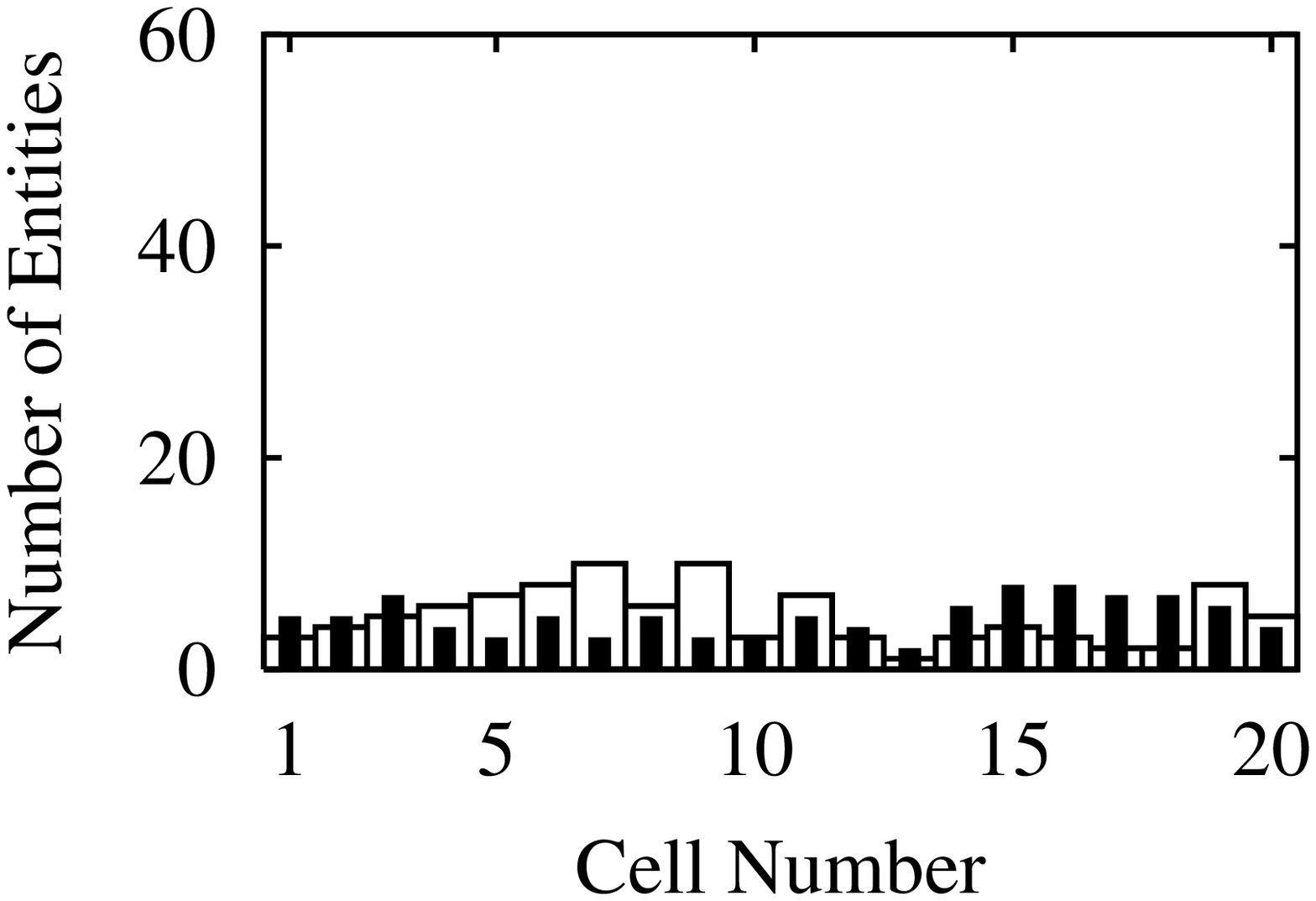}}
{Distributions for 
$\vec{P}=(-1,2,-2,1)$ and 
$D_1=D_2=0.5$ (top), $D_1=50$, $D_2=0.5$ (second), 
$D_1=5000$, $D_2=0.5$ (third), $D_1=5000$, $D_2=5$ (bottom).}
{fig8}

\subsubsection*{\sss Noise-induced self-organization}

At small noise levels $D_a$, 
we will just find more or less 
homogeneous distributions of the entities. This
is already different from the cases of agglomeration, 
segregation, and lane formation we have
discussed before. Self-organization is also not found at higher
noise amplitudes $D_a$, 
as long as we assume that they are the same in both subpopulations
(i.e., $D_1 = D_2$).
However, given that the fluctuation amplitude $D_2$ in subpopulation 2 is
small, we find an agglomeration in subpopulation 2, if the noise level
$D_1$ in subpopulation 1 is medium or high, so that subpopulation 1
remains homogeneously distributed. The order in subpopulation 
2 breaks down, as
soon as we have a relevant (but still small) noise level $D_2$ in
subpopulation 2 (see Fig.~\ref{fig8}). 
\par
Hence, we have a situation
where asymmetric noise with $D_1\ne D_2$ can facilitate self-organization in a
system with completely homogeneous initial conditions and interaction
laws, where we would not have ordering without any noise. We call this 
phenomenon noise-induced self-organization. It is to be distinguished
from the noise-induced increase in the degree of ordering discussed
above, where we have self-organization even without noise, 
if only the initial conditions are not fully homogeneous.
\par
The role of the noise in subpopulation 1 seems to be the following: Despite of
the attractive interaction with subpopulation 2, it 
suppresses an agglomeration in subpopulation 1, 
in particular at the places where subpopulation 2 agglomerates. 
Therefore, the repulsive interaction of
subpopulation 2 with subpopulation 1 is effectively reduced. As a
consequence, the attractive self-interaction within subpopulation 2
dom-\linebreak\newpage
\noindent inates, which gives rise to the observed agglomeration. 
\par
The temporal development of the distribution of entities and of the
overall success in the subpopulations gives additional information
(see Fig.~\ref{fig9}).
As in the case of lane formation, the overall success fluctuates 
strongly, because the subpopulations do not separate from each other,
causing ongoing interactions. Hence, the resulting
distribution is not stable, but changes continuously. It can,
therefore, happen, that clusters of subpopulation 2 merge, which is
associated with an increase of overall success in subpopulation 2
(see Fig.~\ref{fig9}). 
\par\abb{
\vspace*{-5mm}
\hspace*{6mm}\includegraphics[height=6cm, angle=-90]{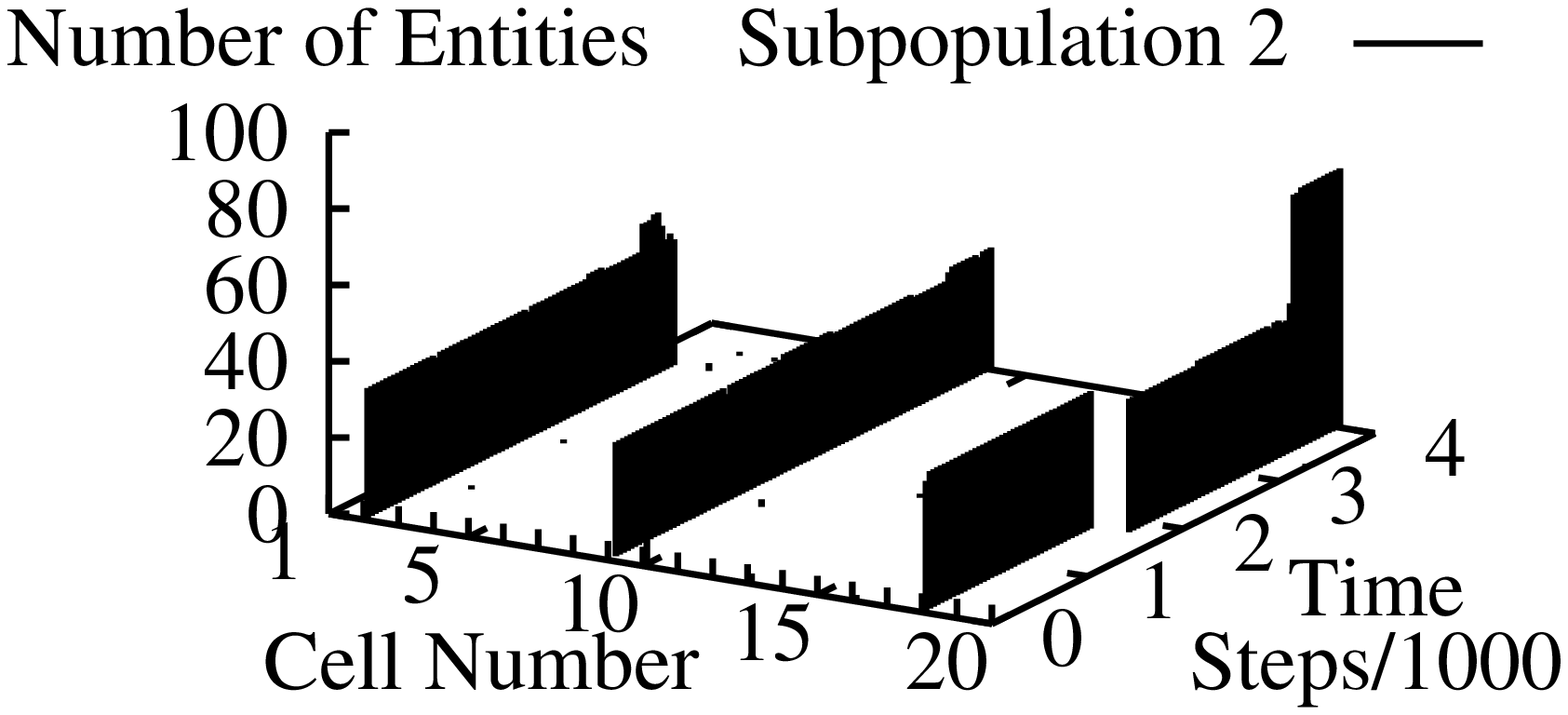}\\[-5mm]
\hspace*{-2mm}\includegraphics[height=5.5cm, angle=-90]{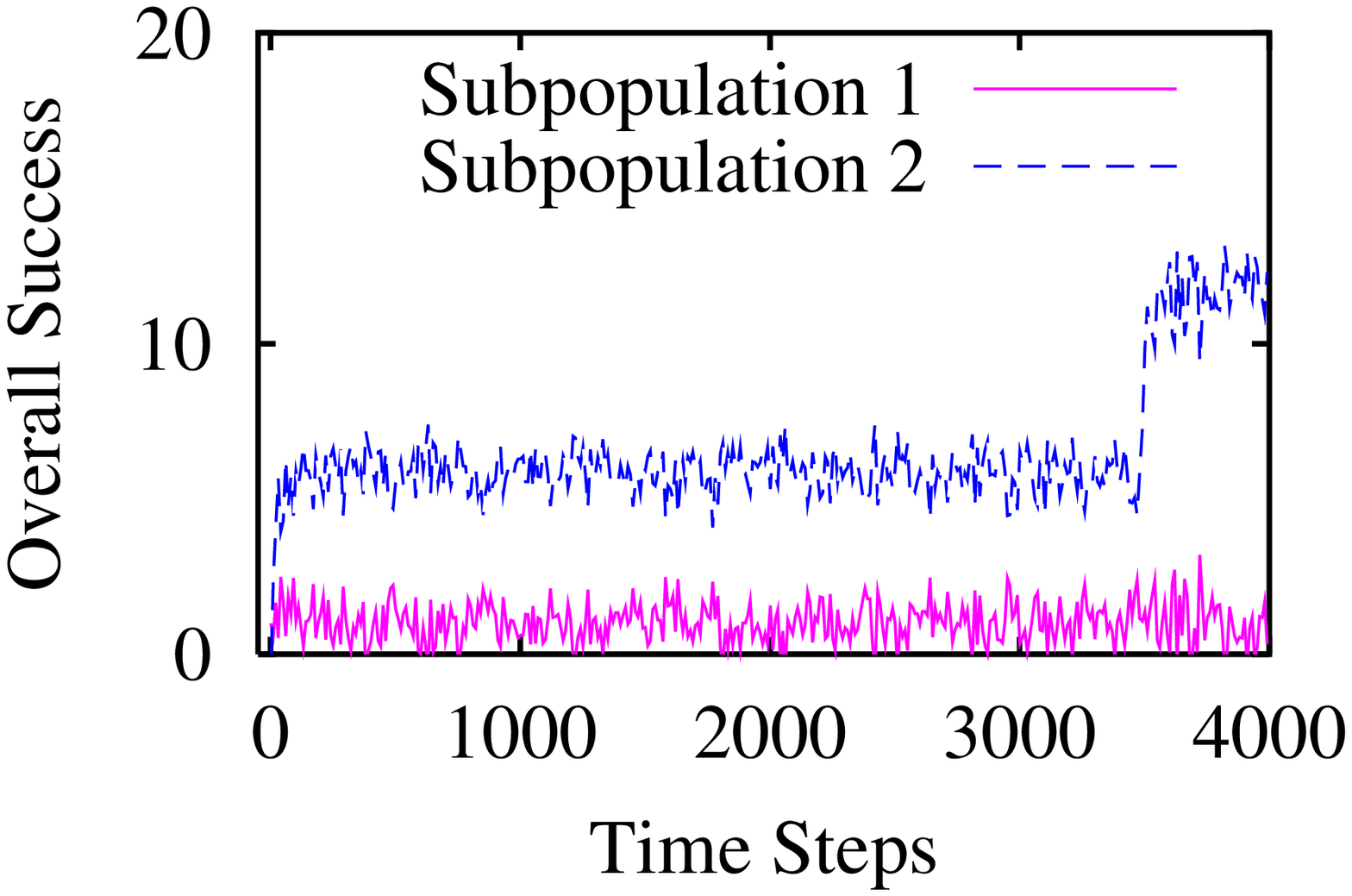}}
{Temporal evolution of the distribution of entitities within subpopulation
$a=2$ (top) and of the overall successes (bottom)
for $\vec{P}=(-1,2,-2,1)$ and $D_1=50$, $D_2=0.5$.}
{fig9}

\section{\large Conclusions}

We have proposed a 
game theoretical model for self-organization in
space, which is applicable to many kinds 
of biological, economic, and social systems with various types of 
profitable or competitive self- and cross-interactions
between subpopulations of the system. Depending on the structure of
the payoff matrix, we found several different self-organization
phenomena like agglomeration, segregation, or lane formation. It
turned out that medium noise strengths can increase the resulting
level of order, while a high noise level leads to more or less homogeneous
distributions of entities over the available space. The mechanism of
noise-induced ordering in the above discussed systems with
short-range interactions seems to be the following: Noise extends a
``pre-ordering'' phase by keeping up a quasi-continuous distribution 
of entities, which allows a long-range ordering. For asymmetric
payoff matrices, we can even have the phenomenon of noise-induced
self-organization, although we start with completely homogeneous
distributions and homogeneous (translation-invariant) payoffs.
However, the phenomenon requires different noise amplitudes in both
subpopulations. The role of noise is to suppress agglomeration in one
of the subpopulations, in this way reducing repulsive effects that
would suppress agglomeration in the other subpopulation.
\par
We point out that all the above results can be semi-quantitatively understood 
by means of a linear stability analysis of a related continuous
version of the model \cite{helvic}. This continuous version indicates
that the linearly most unstable modes are the ones with the shortest
wave length, so that\linebreak\newpage 
\noindent one does not expect a characteristic length scale 
in the system. This is different from reaction-diffusion systems, 
where the most unstable mode has a finite wave length, which gives
rise to the formation of periodic patterns. Nevertheless, the
structures evolving in our model are spatially extended, but
non-periodic. The spatial extension is increasing with the fluctuation 
strength, unless a critical noise amplitude is exceeded.
\par
For a better agreement with real systems,
the model can be generalized in many ways. 
The entities may perform a biased or unbiased random walk in space. 
One can allow random jumps to neigboring cells 
with some prescribed probability. 
This probability may depend on the subpopulation, and thus we can
imitate different mobilities of the considered subpopulations.
Evolution is slowed down by introducing a threshold, fixed or random,
so that the entities change to other
cells only if the differences in the relevant 
successes are bigger than the imposed threshold. 
The model can be also generalized to higher dimensions, 
with expected interesting patterns of self-organized structures.   
\par
In general, the random variables $\xi_\alpha(t)$ in the definition of
the success functions can be allowed to have different variances
for the considered cell $i$ and the 
neighboring cells, with the interpretation that the uncertainty 
in the evaluation of the success in the 
considered cell is different (e.g. smaller) 
than that in the neighboring cells. 
Moreover, the uncertainities can be different for various
subpopulations, which could reflect to some extent their different
knowledge and 
behavior. 
\par
One can as well study systems with more than two subpopulations, 
the influence of long-range interactions, etc. The entities 
can also be allowed to jump to more remote cells.
As an example, the following update rule could be implemented:
Move entity $\alpha$ from cell $i$ to the cell
$(i+l)$ for which
\begin{equation}
S_a^{\prime\prime}(i+l,t) = d^{|l|} c(i+l,t) S_a(i+l,t) 
\end{equation}
is maximal  ($|l| = 0,1,...,l_{\rm max})$. 
If there are $m$ cells in the range $\{(x-l_{\rm
max}),\dots,(x+l_{\rm max})\}$
with the same maximal value, choose one of them randomly with probability $%
1/m$. According to this, when spontaneously moving to another cell, the 
entity prefers cells in the neighborhood with higher success. The indirect
interaction behind this transition, which is based on the observation or
estimation of the success in the neighborhood, is short-ranged if
$l_{\rm max} \ll I$, otherwise long-ranged. Herein, $l_{\rm max}$ 
denotes the maximum number of cells which an entity
can move within one time step. 
The factor containing $d$ with $0 < d < 1$ allows to
consider that it is less likely to move for large distances, if this is not
motivated by a higher success. A value $d<1$ may also reflect the fact that
the observation or estimation of the success over large distances becomes
more difficult and less reliable.\\[5mm]
{\em Acknowledgments:}
D.H. thanks E\"ors Szathm\'{a}ry and
Tam\'{a}s Vicsek for inspiring discussions and
the German Research Foundation (DFG) 
for financial support by a Heisenberg scholarship.
T.P. is grateful to the Alexander-von-Humboldt Foundation 
for financial support during his stay in Stuttgart.

\begin{footnotesize}

\end{footnotesize}
\end{document}